\documentclass[aps,twocolumn,showpacs,prb,superscriptaddress,floatfix,10pt]{revtex4-1}

\usepackage{graphicx,amsmath, amssymb, amsthm, amsfonts, mathtools}
\usepackage[usenames,dvipsnames]{xcolor}
\usepackage{float}
\usepackage{dsfont,soul}

\usepackage{hyperref}			   
\usepackage{braket}

\usepackage{mathrsfs}

\def\beq{\begin{equation}}
\def\eeq{\end{equation}}
\def\beqa{\begin{eqnarray}}
\def\eeqa{\end{eqnarray}}
\def\bsub{\begin{subequations} \begin{align}}
\def\esub{\end{subequations} \end{align}}

\def\da{\downarrow}
\def\ua{\uparrow}

\newcommand{\be}{\begin{equation}}
\newcommand{\ee}{\end{equation}}

\newcommand{\rmd}{{\rm{d}}}
\newcommand{\rme}[1]{{\rm{e}}^{#1}}
\newcommand{\matb}{\left(\begin{array}}
\newcommand{\mate}{\end{array}\right)}
\newcommand{\av}[1]{\lan #1 \ran}
\newcommand{\lan}{\left\langle}
\newcommand{\ran}{\right\rangle}
\newcommand{\mand}{\quad\text{ and }\quad}
\newcommand{\ha}{\frac{1}{2}}
\newcommand{\comma}{\quad , \quad}
\newcommand{\tildel}{\tilde{\delta}}
\newcommand{\lt}{\left(}
\newcommand{\rt}{\right)}
\newcommand{\lqq}{\left[}
\newcommand{\rqq}{\right]}

\begin{document}

\title{Disorder enhanced quantum many-body scars in Hilbert crystals}

\author{Bart van Voorden} 
\affiliation{Institute for Theoretical Physics, University of Amsterdam, Science Park 904, 1098 XH Amsterdam, the Netherlands}

\author{Matteo Marcuzzi} 
\affiliation{School of Physics and Astronomy, University of Nottingham, Nottingham, NG7 2RD, United Kingdom}

\author{Kareljan Schoutens} 
\affiliation{Institute for Theoretical Physics, University of Amsterdam, Science Park 904, 1098 XH Amsterdam, the Netherlands}
\affiliation{QuSoft, Science Park 123, 1098 XG Amsterdam, the Netherlands}

\author{Ji\v{r}\'{i} Min\'{a}\v{r}} 
\affiliation{Institute for Theoretical Physics, University of Amsterdam, Science Park 904, 1098 XH Amsterdam, the Netherlands}
\affiliation{QuSoft, Science Park 123, 1098 XG Amsterdam, the Netherlands}

\date{\today}


\begin{abstract}	

We consider a model arising in facilitated Rydberg chains with positional disorder which features a Hilbert space with the topology of a $d$-dimensional hypercube. This allows for a straightforward interpretation of the many-body dynamics in terms of a single particle one on the Hilbert space and provides an explicit link between the many-body and single particle scars. 
Exploiting this perspective, we show that an integrability-breaking disorder enhances the scars followed by inhibition of the dynamics due to strong localization of the eigenstates in the large disorder limit. Next, mapping the model to the spin-1/2 XX Heisenberg chain offers a simple geometrical perspective on the recently proposed Onsager scars [PRL {\bf 124}, 180604 (2020)], which can be identified with the scars on the edge of the Hilbert space. This makes apparent the origin of their insensitivity to certain types of disorder perturbations.

\end{abstract}

\maketitle

\emph{Introduction. }
The understanding of thermalization and relaxation dynamics is at the forefront of research on quantum many-body systems out-of-equilibrium. 
Since the formulation of the eigenstate thermalization hypothesis~\cite{Deutsch1991, Srednicki1994,Rigol2008}, predicting fast thermalization following a quench from most many-body states, many exceptions to this behaviour have been identified. The prominent examples are integrable~\cite{Sutherland2004,Takahashi_2005_Book} and many-body localized (MBL) systems~\cite{Gornyi2005, Basko2006,Serbyn_2013_PRL,Huse_2014_PRB, Nandkishore2014,Imbrie_2016_JStatPhys,Imbrie_2017_AnnPhys, Alet_2018, Abanin_2019_RMP}. 
A recently added category are quantum many-body scars (QMBS)~\cite{Turner2018, Turner2018a}, which are particular eigenstates responsible for slow decay and oscillatory behaviour of observables following a quantum quench from certain initial states, typically close to a product state, as observed in Ref.~\cite{Bernien2017} realizing the so-called PXP model~\cite{Lesanovsky_2012_PRA}.
This has triggered a great interest in QMBS in settings ranging from constrained to driven
\cite{Choi,Lin2018,Iadecola2019,Surace_2019,Lin_2019,Bull_2020,Mark_2019,Lin_2020,Yang_2020_PRL,
Ho2018a, Bull2019,Pancotti_2019,Roy_2019_PRR,
Ok2019,
Khemani2018,
Hallam_2019_NatComm, Jansen_2019_PRB, Moudgalya_2019_PRB, Wilming_2019_PRL, Michailidis_2020_PRX,Andreev_2019,Werman_2020,
DeTomasi_2019_PRB, Khemani_2019, Sala_2020_PRX,Karpov_2020,
Moudgalya_2018_PRB,Moudgalya2018,Shiraishi_2019_JStatMech,Mark_2020,Moudgalya_2020,
Chattopadhyay2019, Schecter2019, Iadecola_2019_PRL,Shibata_2019,
Lee_2020,
Chen_2020_PRA,
Sinha_2019,Villasenor_2020,
Pai2019, Pizzi_2019,Sugiura_2019,Zhao2020,Mukherjee_2020,Jen_2020_PRR,Fan_2020_PRA,
Yang_2020_PRL, Michailidis_2020_PRX}
and recently also disordered systems \cite{Mondragon_2020}.

QMBS owe their name to the single-particle quantum scars \cite{Kaplan_1998_AnnPhys, Kaplan_1999} which were in turn inspired by particle motion in classical billiards. In both the quantum and classical cases, it is the shape of the billiard boundary, such as the celebrated Bunimovich stadium or cardioid shape \cite{Bunimovich_1979, Robnik_1983_JPhysA}, which causes the motion of the particle to be generically ergodic. The exception to this rule is a set of periodic trajectories, around which the density of certain wavefunctions - the scars - is enhanced in the quantum case.

Here we analyze a model of spins-1/2, which describes a chain of Rydberg atoms with open boundaries under a facilitation condition \cite{Marcuzzi_2017_PRL}.
Representing the Hilbert space as a graph, we show that it corresponds to a truncated hypercube with the dimension given by the number of spin clusters (cf. below for definition). 

This allows us to identify the QMBS as single particle scars on the Hilbert space.
Building on the graph representation of the Hilbert space, an approach also exploited in the studies of MBL \cite{Altshuler_1997_PRL, Gornyi2005, Basko2006, Monthus_2010_PRB, deLuca_2013, Luitz_2015_PRB, Pietracaprina_2016_PRB, Ghosh_2019_PRB, Roy_2019_PRR}, we demonstrate that the scar signatures are \emph{enhanced} in the presence of disorder, naturally emerging from the positional disorder of the atoms.
Finally, exploiting the mapping of the present model to the Heisenberg spin-1/2 XX chain \cite{Ostmann_2019_PRA}, we identify the recently proposed Onsager scars \cite{Shibata_2020_PRL,Vernier_2019_JStatMech} with scars corresponding to sparse eigenstates residing at the ``edge'' of the Hilbert space.
This provides intriguing connections between QMBS and single-particle scars and highlights the utility of a graph-theoretical approach to many-body dynamics, which has been advocated also in the studies of quantum chaos \cite{Kottos_1999_AnnPhys,Schanz_2000_PRL,Smilansky_2007_JPhysA, Smilansky_2013, Lucas_2019}, integrability \cite{Chapman_2020_Quantum}, QMBS \cite{Hudomal_2020_CommPhys} and fermionic and exchange models \cite{Decamp_2020_PRR_0, Decamp_2020_PRR}.

\emph{The model. }
We consider a one dimensional chain of $M$ Rydberg atoms along the $z$-axis, with open boundaries and spaced by $r_0$. We denote the ground and excited (Rydberg) state as $\left| \downarrow\right\rangle$, $\left| \uparrow\right\rangle$. The corresponding Hamiltonian reads
\begin{equation}
	\label{eq:HRy}
	H_{\rm Ry} = \sum_k \frac{\Omega}{2}\sigma_k^x + \Delta n_k + \sum_{l>k}V(|{\bf r}_k-{\bf r}_l|) \, n_k n_l ,
\end{equation}
where $\sigma_k^x = \left|\uparrow_k \right\rangle\left\langle\downarrow_k \right|+\left| \downarrow_k\right\rangle\left\langle\uparrow_k \right|$, $n_k = \left|\uparrow_k \right\rangle\left\langle \uparrow_k\right|$, and $V(r) = C_\alpha/r^\alpha$, $r=|{\bf r}|$. $C_\alpha$, which we take to be positive, is the interaction strength coefficient with $\alpha=3\,(6)$ for dipole-dipole (Van der Waals) interaction. The positions of the atoms are ${\bf r}_k = (0,0,(k-1)r_0)+\delta {\bf r}_k$, where $\delta {\bf r}_k$ describes the disorder which induces the disorder in energy. Denoting $V_{\rm NN}=V(r_0)$ and $V_{\rm NNN}=V(2 r_0)$, we define an energy shift for a pair of nearest neighbours $\delta V_k = V_{\rm NN}-V(|{\bf r}_{k+1}-{\bf r}_k|)$.

It has been shown in \cite{Marcuzzi_2017_PRL} that under the facilitation condition $\Delta = -V_{\rm NN}$ and in the regime $V_{\rm NN} \gg \Omega, \delta V_k$ the Hamiltonian (\ref{eq:HRy}) effectively reduces to
\beq
	H_{\rm eff}=\Delta N_{\rm cl} + \sum_k \frac{\Omega}{2} \sigma^x_k P_{\braket{k}} + \delta V_k n_k n_{k+1} + V_{\rm NNN} n_k n_{k+2},
	\label{eq:Heff}
\eeq
where $P_{\braket{k}}=n_{k-1}+n_{k+1}-2 n_{k-1} n_{k+1}$ and $N_{\rm cl}=\sum_{k} n_{k}(1-n_{k+1})$, $n_0=n_{N+1}=0$, denotes the number of clusters, which are blocks of consecutive spin excitations (e.g. the configuration $\downarrow \downarrow \boxed{\uparrow \uparrow} \downarrow \boxed{\uparrow \uparrow \uparrow}$ contains two clusters highlighted by boxes). The projector $P_{\braket{k}}$ ensures the clusters cannot merge nor disappear and hence their number represents a conserved charge, $[N_{\rm cl},H_{\rm eff}]=0$. For each $N_{\rm cl}$, the topology of the Hilbert subspace of (\ref{eq:Heff}) is that of a truncated hypercube of dimension $d=2 N_{\rm cl}$ \cite{Suppl}. 

\begin{figure}[t!]
	\centering
	\includegraphics[width=0.45\textwidth]{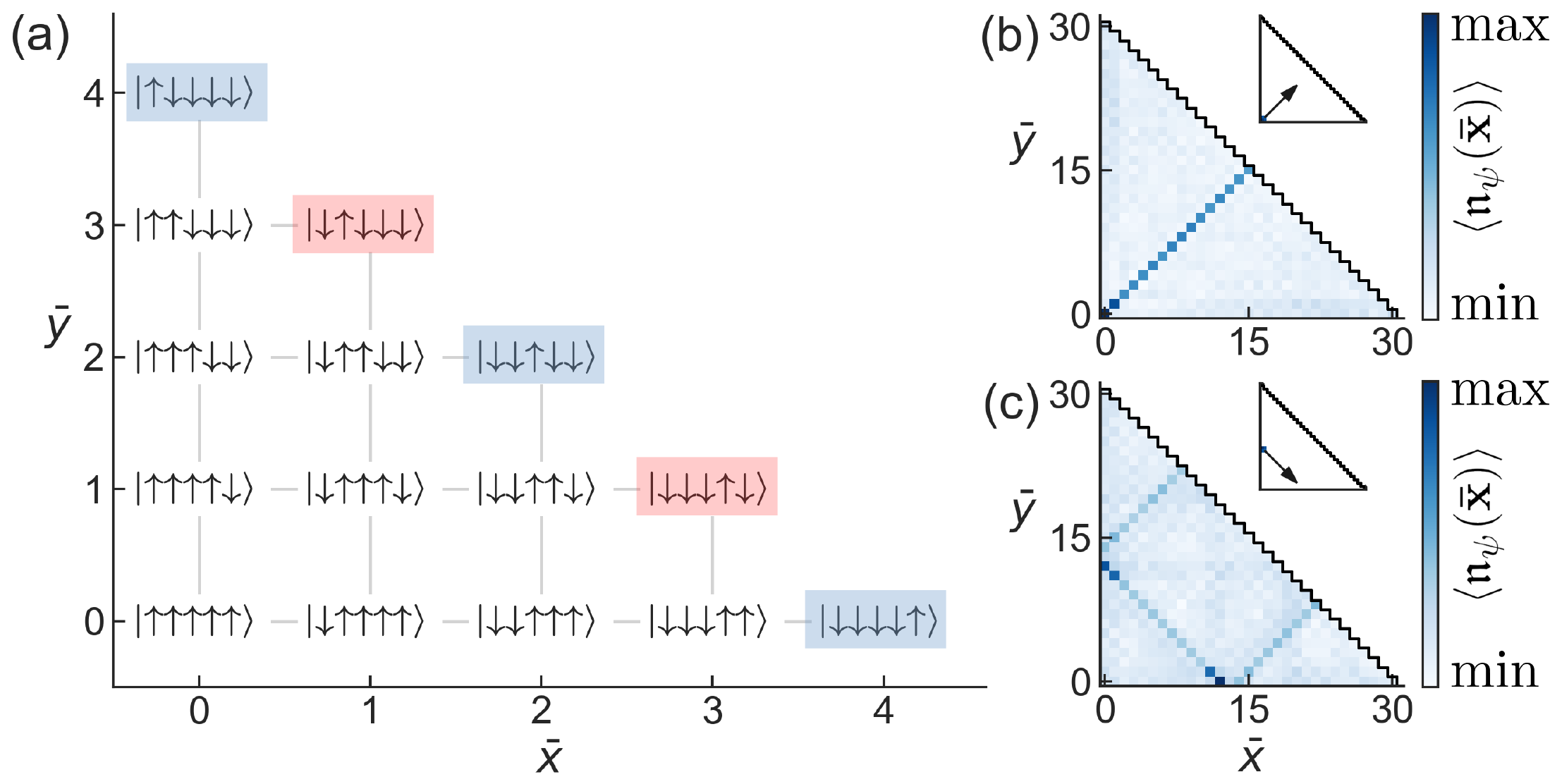}
	\caption{
		{\bf (a)} Hilbert space structure for $M=5$ in the $N_{\rm cl}=1$ sector. The (blue, red) boxes highlight the respective phases (-,+) of basis states constituting a specific sparse eigenvector (a scar).
		{\bf (b,c)}
		The occupation Eq.~(\ref{eq:triangle_localautocor}) with $(\Omega \tau_0, \Omega \tau_1)=(175,700)$ for a quench from the initial state Eq.~(\ref{eq:Gauss}) with $w=2$ and the initial momenta ${\bf p}$ and positions $ \bar{\bf x}_0$ indicated in the insets.			
	}	
	\label{fig:1}
\end{figure}

In what follows we will be particularly focusing on the $N_{\rm cl}=1$ sector for which the Hilbert space can be represented as a square lattice with a triangular boundary. Each site $(\bar{x},\bar{y})$ of this lattice corresponds to a state
\beq
	\label{eq:bstate}
	\ket{{\bf \bar{x}}} \equiv \ket{\bar{x},\bar{y}} = \ket{[\da]_{\bar{x}} \ua \ldots \ua [\da]_{\bar{y}}},
\eeq
Here, $[\da]_\ell$ labels a string of consecutive down spins of length $\ell$. The boundaries are determined by the natural conditions $x \geq 0$, $y \geq 0$ and $x + y < M$, cf. Fig. \ref{fig:1}a.
$H_{\rm eff}$ projected on the $N_{\rm cl}=1$ sector can be written as
\begin{subequations}
\label{eq:single_ham}
	\begin{align}
	H &= H_0 + H_{\rm pot} + H_{\rm dis} \label{eq:H} \\	
	H_0 &= \frac{\Omega}{2}\sum_{\bar{\bf x} \in {\cal H}\setminus {\rm b}} \left|\bar{\bf x} \right\rangle \left( \left\langle \bar{\bf x}+1_{\bar{x}} \right| + \left\langle \bar{\bf x}+1_{\bar{y}}\right| \right) + \text{H.c.} \label{eq:H0}\\
	H_{\rm pot} &= V_{\rm NNN} \sum_{\bar{\bf x} \in {\cal H}} \max(0,M-2-(\bar{x}+\bar{y})) \left|\bar{\bf x} \right\rangle\left\langle \bar{\bf x} \right| \label{eq:Hpot} \\	
	H_{\rm dis} &= \sum_{\bar{\bf x} \in {\cal H}} \ket{\bar{\bf x}} \bra{\bar{\bf x}} \delta V_{\bar{\bf x}}	
	\end{align}	
\end{subequations}
where $1_{\bar{x},\bar{y}}$ are unit vectors in the direction $\bar{x},\bar{y}$, ${\cal H}=\{\ket{{\bf \bar{x}}} \, | \, 0 \leq (\bar{x},\bar{y}) < M \wedge \bar{x} + \bar{y} < M \}$, ${\rm b} = \{ \ket{{\bf \bar{x}}} \, | \, \bar{x}+\bar{y}=M-1 \}$ and $\delta V_{\bar{\bf x}}$ is specified in Eq.~(\ref{eq:delta_V}).

$H_0$ can be solved exactly \cite{Suppl} with eigenenergies
\begin{equation}
\label{eq:triangle_energy}
	2 \Omega^{-1} E_{m,n} = 2 \cos\left(\frac{m\pi}{M+2}\right) + 2 \cos\left(\frac{n\pi}{M+2}\right)
\end{equation}
and eigenvectors
\begin{equation}\label{eq:eigenvecs_triangle}
	\left| w_{m,n}\right\rangle  = \sum_{{\sf x},{\sf y}} \left(u_m({\sf x})u_n(-{\sf y}) - u_m(-{\sf y})u_n({\sf x})\right)\left|{\sf x},{\sf y} \right\rangle
\end{equation}
where
\beq
\label{um(x)}
	u_m({\sf x})= \sqrt{\frac{2}{M+2}} \sin \left( \frac{\pi m}{M+2}({\bar{x}+1}) \right)
\eeq
with $m,n \in \{1,2,\ldots,M+1\}$, $m>n$, ${\sf x} = \bar{x}-M/2$ and ${\bf \bar{x}} \in {\cal H}$. All energies are non-degenerate, except for $\left \lceil M/2 \right \rceil$ zero-energy states for which $m+n = M+2$.
It can be shown that the zero-energy subspace is spanned by eigenvectors, which are sparse in the basis Eq.~(\ref{eq:bstate}) \cite{Suppl}. Due to its simple structure, they can be identified as scars in the Hilbert space, cf. Fig.~\ref{fig:1}a. Consequently, one can directly apply the single-particle perspective used in quantum scars on discrete lattices \cite{Fernandez_2014_NJP}. 
In what follows we examine the dynamics following a quantum quench. Motivated by the use of Gaussian wavepackets as probes for single particle scars \cite{Kaplan_1998_AnnPhys, Kaplan_1999, Kaplan_1999_PRE, Wisniacki_2000_PRE}, we introduce effective ``Gaussian'' initial states defined as (up to normalization)
\beq
	\label{eq:Gauss}
	\ket{\psi_{{\bf \bar{x}}_0}^{{\bf p},w}(t=0)} \propto {\cal P} \sum_{\bar{\bf x}} {\rm e}^{-\frac{(\bar{\bf x}-\bar{\bf x}_0)^2}{2 w^2}} {\rm e}^{-i {\bf p} \cdot {\bf \bar{x}}} \ket{\bar{\bf x}}
\eeq
where ${\bf p}=(p_x,p_y)$ are the phases specifying the initial direction of propagation of the ``wavepacket'' and for simplicity we project the state by ${\cal P}$ on four basis states with maximal weight. For future convenience, we define $\ket{\psi_G} \equiv \ket{\psi_{\bar{\bf x}_0 = (0,0)}^{{\bf p}=(\pi/2,\pi/2),w=2}}$. We also define the time-averaged occupation of the basis states in the Hilbert space as
\begin{equation}
	\label{eq:triangle_localautocor}
	\braket{{\frak n}_{\psi}({\bf \bar{x}})}= \frac{1}{\tau_1-\tau_0}\int_{\tau_0}^{\tau_1} {\rm d}t \, \left|\left\langle {\bf \bar{x}} \middle|\psi(t) \right\rangle\right|^2,
\end{equation}
where $\ket{\psi(t)}$ is the time evolved initial state.

%
In Fig.~\ref{fig:1}b,c we show $\braket{{\frak n}_{\psi}({\bf \bar{x}})}$ for different initial states Eq.~(\ref{eq:Gauss}).
It is apparent that the the occupation clearly reveals the scar behaviour in the Hilbert space in exact analogy to the single-particle case.
\begin{figure}[t!]
	\centering
	\includegraphics[width=0.45\textwidth]{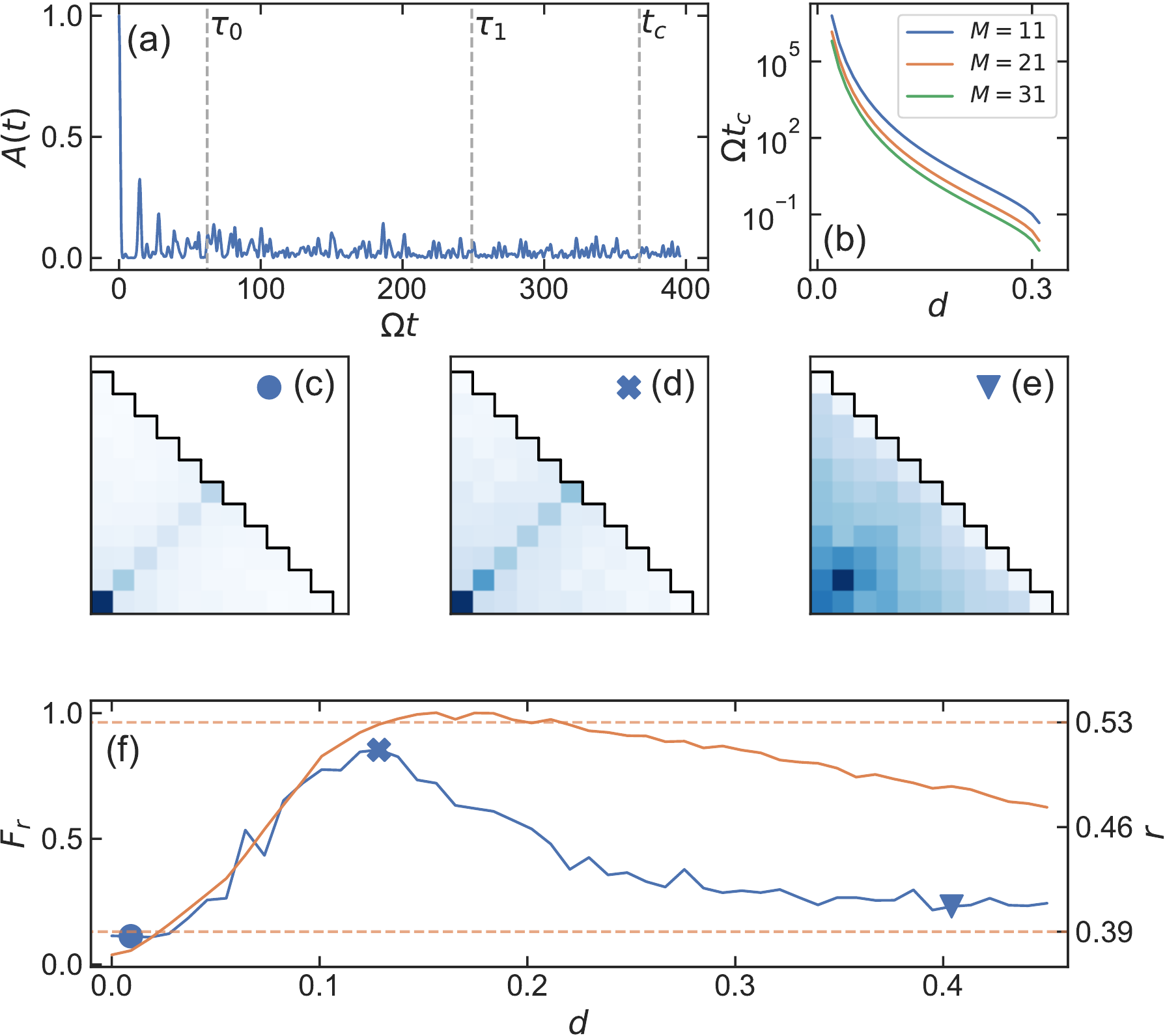}
	\caption{
		{\bf(a)} An example of the autocorrelation $A(t)=|\braket{\psi_G|\psi(t)}|^2$ for $M=11$.
		{\bf (b)} The threshold time $t_c$ vs. disorder strength for various system sizes $M$.
		{\bf (c-e)} Examples of the occupation Eq.~(\ref{eq:triangle_localautocor}) for various disorder strengths indicated by circle, cross and triangle respectively in pane (f).
		{\bf (f)} $F_r$ (blue) and the $r$-statistics (orange) vs. disorder strength $d$ (here $\alpha=6$, $s_0=0.03,\epsilon=9$, $(\Omega \tau_0, \Omega \tau_1)=(50,250)$ and $V_{\rm NN}/\Omega=4$).
	}	
	\label{fig:2}
\end{figure}
%

\emph{Disorder.} Since $H_0$ is integrable, a natural way to break the integrability is provided by positional disorder of the atoms. Denoting $\delta\mathbf{r}_k = (x_k,y_k,z_k)$, the initial position of the $k$-th atom is drawn from a Gaussian probability distribution
$p(\delta\mathbf{r}_k) = (2\pi)^{-3/2}(\prod_{\nu=x,y,z}\sigma_\nu)^{-1} {\rm exp}\left[- \sum_{\nu=x,y,z} \frac{\nu_k^2}{2 \sigma_\nu^2}\right]$ \cite{Marcuzzi_2017_PRL, Ostmann_2019_QSciTech, Tamura_2020_PRA}.

While the primary focus of this article is the analysis of the model (\ref{eq:Heff}),(\ref{eq:single_ham}), to provide a description applicable to a realistic experimental realization, the time dependence of the atom motion ${\bf r}_k(t)$ has to be taken into account. To set up the stage a few remarks are in order. 

First, we consider both the ground and the Rydberg states to be subject to the same harmonic trapping potential $H_{\rm tr}=\sum_k \sum_{\nu=x,y,z} m \omega_\nu^2 \nu_k^2/2$  \cite{Wilson_2019}, where $\omega_\nu$ are the trap frequencies which determine, together with the inverse temperature $\beta=1/k_{\rm B} T$, the disorder through $\sigma_\nu = \sqrt{1/(\beta m \omega_\nu^2)}$ and $m$ is the atom mass.
%
We parametrize the trap frequencies as ${\bf \omega} = (\epsilon^{-1},1,1) \omega_0/d$ which leads to the dimensionless disorder ${\bf s} = (s_x,s_y,s_z)\equiv (\epsilon,1,1)d s_0$, where $s_0=\sigma_0/r_0$ for some $\sigma_0$ and motivated by \cite{Marcuzzi_2017_PRL} we choose $s_0 = 0.03$. Here $\epsilon$ and $d$ tune the shape and the overall strength of the trapping potential where typically $\epsilon > 1$ in a tweezer experiment \cite{Marcuzzi_2017_PRL,Ostmann_2019_QSciTech,Ostmann_2019_PRA}.

Second, we note that the interaction $V(|{\bf r}_k - {\bf r}_l|)$ leads to dynamics entangling the motional and internal degrees of freedom necessitating a fully quantum treatment. This is a difficult problem limiting the applicability of methods  such as exact diagonalization to few sites and small phonon number \cite{Gambetta_2020_PRL}. To proceed, we treat the atomic motion ${\bf r}_k(t)$ as that of a classical particle in a harmonic potential with coordinates $\nu_k(t) = C_{\nu,k} \cos(\omega_\nu t + \phi_{\nu,k})$, where 
$C_{\nu,k} = \sqrt{\nu_k(0)^2 + (q_{\nu,k}(0)/m)^2 / \omega_\nu^2)}$
and
$\phi_{\nu,k} = \arccos\left(\nu_k(0) / C_{\nu,k}\right)$
which are fully specified by the initial position $\nu_k(0) = \delta r_{k,\nu}(t=0)$ and momentum $q_{\nu,k}(0)$. Here, the latter is drawn from an isotropic Boltzmann distribution $p(q_{\nu,k}) \propto {\rm exp}(-\beta q_{\nu,k}^2/(2 m))$.

The third and final comment is that for $V \propto 1/r^\alpha$, the distribution $p(\delta {\bf r}_k)$ leads to the energy probability distribution $p(\delta V)$ with undefined moments, a consequence of rare events when two atoms come arbitrarily close to each other \cite{Ostmann_2019_QSciTech}. This is an artefact, not expected to occur under realistic experimental conditions, of the algebraic form of $V$. For this reason and in order to gain an analytical control, we use a small-displacement approximation
\beqa
\label{eq:delta_V}
	&&\delta V_{\bar{\bf x}} = \sum_{k=\bar{x}+1}^{M-\bar{y}-1} \left[ \frac{C_\alpha}{|{\bf r}_{k+1} - {\bf r}_{k}|^\alpha} -V_{\rm NN} \right] \nonumber \\
			&&\approx -\sum_{k=\bar{x}+1}^{M-\bar{y}-1} \alpha V_{\rm NN} \left[\tilde{\delta}_{z,k} + \frac{1}{2}\left( \tilde{\delta}_{x,k}^2 + \tilde{\delta}_{y,k}^2 -(1+\alpha)\tilde{\delta}_{z,k}^2 \right) \right], \nonumber \\
\eeqa
where $\tilde{\delta}_{\nu,k} = (\nu_{k+1}-\nu_{k})/r_0$. In order to get the occupation (\ref{eq:triangle_localautocor}) with the time-dependent Hamiltonian (\ref{eq:H}) we solve the corresponding Schr\"{o}dinger equation for the wavefunction. In particular we are interested in the properties of the occupation as a function of the disorder. The results for $\ket{\psi(0)}=\ket{\psi_G}$ are shown in Fig.~\ref{fig:2}a,f with examples of $\braket{{\frak n}_{\psi_G}(\bar{\bf x})}$ for three different values of disorder shown in Fig.~\ref{fig:2}c-e. 
The solid blue line in Fig.~\ref{fig:2}f corresponds to a quantity $F_r$ which characterizes the overlap of the occupation with the occupation $\braket{{\frak n}_{\psi}(\bar{\bf x})}_0$ generated by the idealized Hamiltonian $H_0$, Eq.~(\ref{eq:H0}). 
It is defined as $F_r = (F - F_{u})/(1 - F_{u} )$, where $F = \sum_{\bar{\bf x}} \braket{\braket{\tilde{{\frak n}}_{\psi}(\bar{\bf x})}} \braket{\braket{\tilde{{\frak n}}_{\psi}(\bar{\bf x})}}_0$, $F_u$ is given by $F$ with the replacement $\braket{\tilde{{\frak n}}_{\psi}(\bar{\bf x})} \rightarrow \sqrt{2/(M+1)M}$, the tilde denotes the occupations normalized as $\sum_{\bar{\bf x}} \braket{\braket{\tilde{{\frak n}}_{\psi}(\bar{\bf x})}}^2=1$ and the double brackets denote the averaging over disorder realizations (initial conditions).
The rationale behind $F_r$ is that $F_r=1$ when the occupation is that of the idealized scenario of Fig.~\ref{fig:1}b and $F_r=0$ for a featureless uniform occupation. For comparison, the orange solid line shows the level statistics 
$r = \braket{\braket{ \frac{\min(\Delta E_i, \Delta E_{i+1}) }{ \max(\Delta E_i, \Delta E_{i+1}) } }}$ taking the initial conditions, i.e. quenched positional disorder, where the average is taken over all energy differences $\Delta E_i = E_i - E_{i-1}$ of adjacent ordered eigenenergies $E_i \geq E_{i-1}$ of $H$. The values $r \approx 0.39, 0.53$ corresponding to the Poisson and Wigner-Dyson statistics are indicated by the horizontal dashed lines. 
It is apparent from Fig.~\ref{fig:2} that increasing the disorder \emph{enhances} the many-body scars appearing in the occupation.
which can be explained in terms of the eigenstate localization: as the disorder is increased from zero, the eigenstates of $H$ become more and more localized on the Hilbert space square lattice. This initially enhances their overlap with the initial state along the scar path.
We observe similar enhancement also for other initial states and values of disorder and discuss quantitatively the energy landscape of the Hilbert space in \cite{Suppl}.

\begin{figure}[t!]
	\centering
	\includegraphics[width=0.45\textwidth]{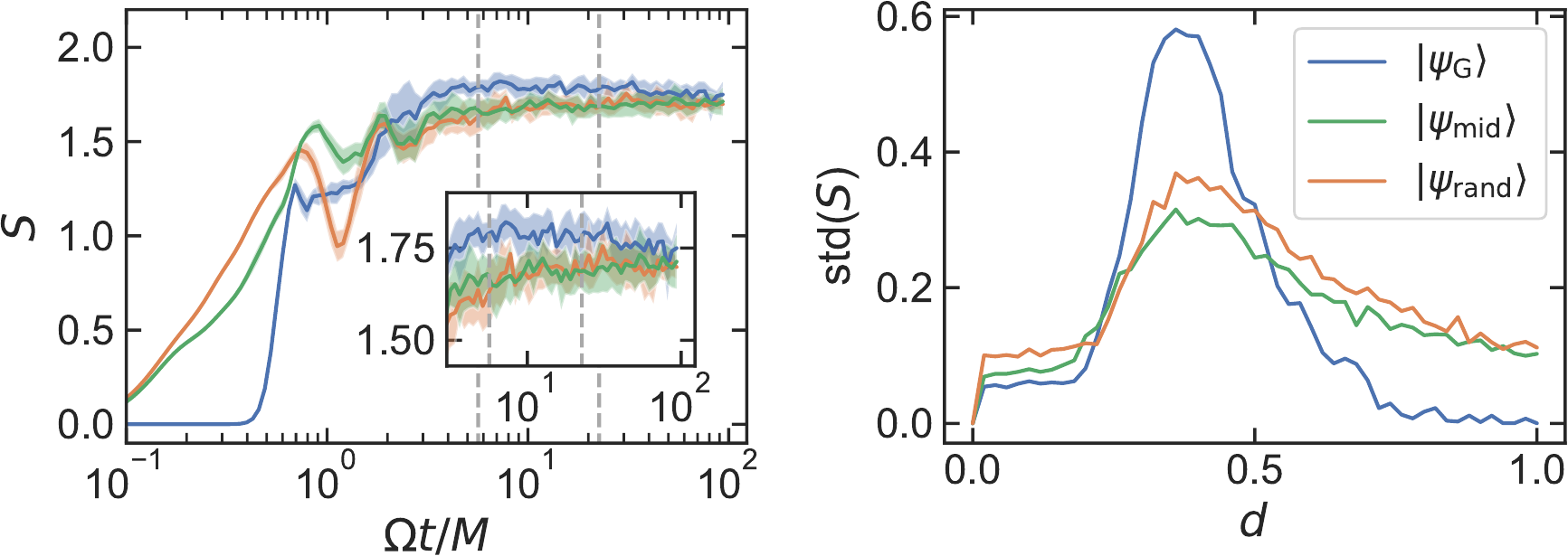}
	\caption{
		{\bf (a)} Evolution of half-chain entanglement entropy $S$ for $d=0.12$ and $M=25$ following a quench from $\ket{\psi_{\rm G}}$ (blue), $\ket{\psi_{\rm rand}}$ (green) and $\ket{\psi_{\rm mid}}$ (orange). The vertical dashed lines indicate the (scaled) times $\tau_0,\tau_1$ used in Fig.~\ref{fig:2} and the inset shows the detail of the late-time evolution.
		{\bf (b)} The standard deviation of the saturated $S$ vs. $d$. Data obtained with 10 realizations of the initial conditions (a) and 300 realizations (b), where static disorder was considered for numerical reasons, yielding a value of the average saturated entropy compatible with (a) within ${\rm std}(S(t \rightarrow \infty))$.
			}	
	\label{fig:3}
\end{figure}

%
\emph{Thermalization.} 
Next we investigate how the scars affect the capacity of the system to thermalise. To this end we consider the time evolution of the (second R\'{e}nyi) entanglement entropy (EE) $S(t) = -{\rm log} \, {\rm Tr}[\rho_A(t)^2]$, 
where $\rho_A(t)$ is the reduced density matrix of subsystem $A$ which we choose to be a half-chain of length $\left \lfloor{\frac{M}{2}}\right \rfloor$. In Fig.~\ref{fig:3}a we plot the time evolution of EE for a quench in the non-integrable regime $d=0.12$ from the Gaussian state $\ket{\psi_{\rm G}}$ (blue), a mid-spectrum eigenstate $\ket{\psi_{\rm mid}}$ of $H$ (orange) and a random state $\ket{\psi_{\rm rand}} \propto \sum_{\bar{\bf x}} c_{\bar{\bf x}} \ket{\bar{\bf x}}$ (green), where $c_{\bar{\bf x}}$ are drawn from a uniform random distribution. Here, $\ket{\psi_{\rm mid}}$ and $\ket{\psi_{\rm rand}}$ are defined on the half-chain so that $S(0)=0$.
%
After the initial rise we observe a slow
growth, cf. \cite{Suppl} for extended discussion, for all the states which we attribute to superscarring, i.e. the fact that \emph{each} basis state either belongs to a scar in the Hilbert space or is adjacent to it. 
%
We also note the initial rise for the Gaussian state happening for $\Omega t \approx M/2$, which corresponds to the geometrical distance from the tip [$\bar{\bf x}=(0,0)$] to the base of the triangular-shaped Hilbert space, cf. Fig.~\ref{fig:1}a.
We note that the scar enhancement is not reflected in the standard deviation of the saturated entropy ${\rm std}(S(t \rightarrow \infty))$ shown in Fig.~\ref{fig:3}b, where the dominant peak around $d \approx 0.3$ corresponds to the transition from non-integrable to integrable as quantified by $r$ \cite{Kjall_2014_PRL, Luitz_2015_PRB, Alet_2018} and hints towards a possible MBL-like phase \cite{Ostmann_2019_PRA}.
%

\emph{Relation to Onsager scars. }
It has been shown in  \cite{Ostmann_2019_PRA} that the spin flip part of $H_{\rm eff}$, Eq.~(\ref{eq:Heff}), can be mapped to the spin-1/2 XX Heisenberg spin chain of length $M+1$
\beq
	\label{eq:HXX}
	\sum_k \sigma^x_k P_{\braket{k}} \rightarrow H_{\rm XX} = \sum_{k=1}^{M} \mu^x_k \mu^x_{k+1} + \mu^y_k \mu^y_{k+1},
\eeq
where $\mu^{x,y,z}$ the Pauli matrices in a $\{\ket{0},\ket{1}\}$ basis. It is related to the $\{\ket{\da}, \ket{\ua}\}$ basis through the mapping $\ua \ua, \da\da \rightarrow 0$, $\ua \da, \da \ua \rightarrow 1$, where the ambiguity is lifted by including fictious boundary spins $(\da)$ to the left and right ends of the chain. Consequently, $\sigma^x_k = \mu^x_k \mu^x_{k+1}, \sigma^y_k = (-1)^{k+1} \prod_{l=1}^{k-1} \mu^z_l \mu^y_k \mu^x_{k+1}, \sigma^z_k = (-1)^{k+1} \prod_{l=1}^{k} \mu^z_l$ and $\delta V$ of Eq.~(\ref{eq:Heff}) maps to non-local disorder given by a string of $\mu^z$ operators \cite{Ostmann_2019_PRA}.

Crucially, the structure of the Hilbert space (connectivity between the basis states) remains unchanged as it is given solely by the spin flip terms \cite{Suppl}. Recently, Ref.~\cite{Shibata_2020_PRL} proposed a class of spin models with $n$ spin components featuring so-called Onsager scars, which are states with perfect revivals of the integrated autocorrelation subject to certain types of integrability-breaking disorder. The simplest instance $n=2$ of this class is $H_{\rm XX}$, Eq.~(\ref{eq:HXX}), with the Onsager scar  $\ket{\psi(\beta)} \propto {\rm exp}[\beta^2 Q^+]\ket{0 \ldots 0} = \sum_{N_{\rm cl}=0}^{\left \lceil (M+1)/2 \right \rceil} \frac{(\beta^2 Q^+)^{N_{\rm cl}}}{N_{\rm cl}!} \ket{0 \ldots 0}$ and $Q^+ = \sum_k (-1)^{k+1} \mu^+_k \mu^+_{k+1}$.
We have intentionally indexed the summation in the definition of $\ket{\psi(\beta)}$ by $N_{\rm cl}$ as each term corresponds to a superposition of $N_{\rm cl}$ pairs $\ket{\ldots 1_k 1_{k+1} \ldots}$, i.e \emph{single} Rydberg spins $\ua$. The projection of $\ket{\psi(\beta)}$ on the $N_{\rm cl}=1$ sector is nothing but the scar indicated in Fig.~\ref{fig:1}a.

This allows for the following identifications: \emph{(i)} The $\left \lceil (M+1)/2 \right \rceil$ eigenstates which form the special band in the plot of the eigenstate's EE, cf. Fig.~2a in \cite{Shibata_2020_PRL}, correspond to different cluster sectors of $H_{\rm eff}$. \emph{(ii)} The projection of $\ket{\psi(\beta)}$ on $N_{\rm cl}=1$ sector is the scar corresponding to the $(0,M-1)-(M-1,0)$ diagonal, i.e. the \emph{edge} of the Hilbert space, cf. Fig.~\ref{fig:1}a, which is comprised only of single Rydberg spin excitations. This interpretation bears to other $N_{\rm cl}$ as well. Furthermore, the simple structure of the Hilbert space allows for a straightforward visualisation of why certain types of the integrability-breaking disorder do not affect the Onsager scars, such as Eq.~(13) in \cite{Shibata_2020_PRL}. Another example naturally realized in the Rydberg systems is the disorder of Eq.~(\ref{eq:Heff}) which affects all but the isolated Rydberg spins.

\emph{Experimental considerations. }
We have simulated the time evolution with the assumption that the atomic trajectories are that of classical particles in a harmonic potential, independent of their internal state. To estimate the effect of the Rydberg interactions on the atomic motion and hence the disorder energies, we consider $\braket{\delta V(n_{\rm NN})}$ to be the expectation value of $\delta V_{\bar{\bf x}}$, Eq.~(\ref{eq:delta_V}), corresponding to basis state $\ket{\bar{\bf x}}$ containing $n_{\rm NN}$ nearest neighbours and evaluated using $p(\delta {\bf r}_k)$.
Analogously, we define $\braket{\delta V(n_{\rm NN})}_{\rm int}$ where the equilibrium positions of the atoms are taken in the presence of the interactions \cite{Suppl}. The difference between the two provides an estimate for a threshold timescale beyond which the atomic motion cannot be treated as independent of the internal state and we define $t_c \equiv 2\pi \hbar/(\braket{\delta V(M-1)}_{\rm int} - \braket{\delta V(M-1)})$. The plot of $t_c$ vs. $d$ is shown in Fig.~\ref{fig:2}b with an example of $t_c$ indicated in Fig.~\ref{fig:2}a.
Thus, for $d \approx 0.1$, the present analysis holds for $\Omega t = O(100)$ for $M$ of few tens, sufficient to capture the behaviour of the time-averaged occupation in a realistic experimental setting.

\emph{Outlook.}
In this work we have highlighted how the structure of the Hilbert space, resembling that of a hypercubic crystal,
provides useful insights in the non-equilibrium dynamics in spin chains. This allowed us to identify quantum many-body scars as single particle scars in the Hilbert space, link them to the Onsager scars and show how their signature is enhanced by disorder.
This provides a number of interesting openings, such as the interpretation of the disordered Heisenberg XXZ spin chain as that of an Anderson model on a hypercubic lattice, which is relevant to the ongoing discussion about the scaling of the Thouless time in many-body systems \cite{Suntajs_2019,Sierant_2020_PRL}. 
It would be also interesting to explore the role of sparse eigenvectors, which play an important role in various applications, such as in the signal analysis of networks \cite{Teke_2017, Teke_2017b}, in the context of many-body Hamiltonians and their graph-theoretic representations \cite{Decamp_2020_PRR,Decamp_2020_PRR_0,Chapman_2020_Quantum, Chen_2019}. 
Finally, to describe the entangling dynamics between the motional and internal degrees of freedom, new approaches, such as the variational ansatz based on non-Gaussian states \cite{Shi_2018_AnnPhys}, need to be investigated.

\emph{Acknowledgments. }
We are very grateful to V. Gritsev, Neil J. Robinson, W. Buijsman, W. Vleeshouvers, A. Urech, V. Alba, Y. Miao and O. Gamayun for fruitful discussions. This work is part of the Delta ITP consortium, a program of the Netherlands Organisation for Scientific Research (NWO) that is funded by the Dutch Ministry of Education, Culture and Science (OCW).
M.M. gratefully acknowledges funding from the University of Nottingham under a Nottingham Research Fellowship Scheme.



%


\clearpage

\onecolumngrid

\begin{center}
{\Large{SUPPLEMENTAL MATERIAL}}
\end{center}

\setcounter{equation}{0}
\setcounter{figure}{0}
\renewcommand{\theequation}{S\arabic{equation}}
\renewcommand{\thefigure}{S\arabic{figure}}
\renewcommand\thesection{\Roman{section}}

\section{Eigenstates of $H_0$}
\label{app:Eigenstates}

In the absence of disorder, the mapping Eq.~(\ref{eq:HXX}) allows for exact solution of the model through the Jordan-Wigner transformation. Nevertheless, the simple structure of the Hilbert space associated with the spin flip Hamiltonian allows for the following more explicit construction. Focusing on $H_0$, Eq.~(\ref{eq:H0}) and Fig.~\ref{fig:1}a, the eigenstates and eigenergies can be found as follows.

We first embed the square lattice with triangular boundaries in a larger lattice with square boundaries, where the side of the square is comprised of $M+1$ sites and the lower left corner of the square has coordinates ${\bar{\bf x}}=(0,0)$. Lets now consider a hopping on an open chain of $M+1$ sites, such that the positions of the sites ${\bar x}=0,\ldots, M$ respectively. It is straightforward to find the eigenfunctions of such hopping Hamiltonian which read
\beq
	u_m({\bar x})=\sqrt{\frac{2}{M+2}} \sin \left( \frac{\pi m}{M+2} (\bar{x}+1) \right)
	\label{eq:u_chain}
\eeq
for $m=1,\ldots,M+1$. This can be understood as that the eigenfunction has to vanish beyond the boundaries of the chain, i.e. for $\bar{x}=-1$ and $\bar{x} = M+2$. The eigenvectors of $H_0$ can then be obtained simply as  a product of the open-chain solutions (\ref{eq:u_chain}) with the extra requirement that the coefficients of the eigenvectors have to vanish on the diagonal ${\bar{\bf x}}: (0,M+1)-(M+1,0)$ of the embedding square. To this end it is convenient to transform to the coordinates symmetric with respect to the centre of the square, namely
\beq
	{\bf x} = ({\sf x},{\sf y}) = \bar{\bf x} - \left( \frac{M}{2} , \frac{M}{2} \right).
	\label{eq:shift}
\eeq
The Eq.~(\ref{eq:u_chain}) becomes
\beq
	u_m({\sf x})=\sqrt{\frac{2}{M+2}} \sin \left( \frac{\pi m}{M+2} {\sf x} + \frac{\pi m}{2} \right) =
	\begin{cases}
		(-1)^{\frac{m-1}{2}}\cos \left( \frac{\pi m {\sf x}}{M+2} \right) \;\; m \; {\rm odd} \\
		(-1)^{\frac{m}{2}}\sin \left( \frac{\pi m {\sf x}}{M+2} \right) \;\; m \; {\rm even}.
	\end{cases}
	\label{eq:u_chain2}
\eeq
The solutions on the Hilbert space of $H_0$ are then obtained by requiring that the wavefunctions vanish on the diagonal ${\sf y}=-{\sf x}$ of the embedding square. This is achieved by the antisymmetrization of the solutions (\ref{eq:u_chain2}) leading to
\beq
	\left| w_{m,n}\right\rangle  = \sum_{{\sf x},{\sf y}} \left(u_m({\sf x})u_n(-{\sf y}) - u_m(-{\sf y})u_n({\sf x})\right)\left|{\sf x},{\sf y} \right\rangle,
	\label{eq:eigenvecs_triangle2}
\eeq
where the solutions of the Hamiltonian $H_0$ are given by, with the help of (\ref{eq:shift}), $\bar{x},\bar{y} \in \{0,M-1\}$, $\bar{x}+\bar{y} < M$ and $m > n$, which yields $M(M+1)/2$ eigenvectors, a number equal to the dimension of $H_0$. The corresponding eigenenergies are
	\begin{equation}\label{eq:triangle_energy2}
	2 \Omega^{-1} E_{m,n} = 2 \cos\left(\frac{m\pi}{M+2}\right) + 2 \cos\left(\frac{n\pi}{M+2}\right) \, ,
	\end{equation}	 
which posses the inversion symmetry $E_{M+2-n, M+2-m} = -E_{m,n}$.

\subsection{Scars as sparse eigenstates}

Given a basis ${\cal H} = \{\ket{b_j}\}$ and a state $\ket{v_i} = \sum_i c_{ij} \ket{b_j}$ we define
\beq
	{\cal S}_{\ket{v_i}} = \{ \ket{b_j} \, | \, \braket{v_i | b_j} \neq 0 \}
\eeq
which is a set of all basis states with non-zero overlap with $\ket{v_i}$. We then define the sparsity of the state $\ket{v_i}$ in the usual sense as
\beq
	\mathscr{S}_{\ket{v_i}} = \frac{|{\cal H}| - |{\cal S}_{\ket{v_i}}|}{|{\cal H}|}.
	\label{eq:sparsity}
\eeq
We note that there are $\left \lceil M/2 \right \rceil$ zero-energy eigenstates $E_{m,n}=0$ for $m+n = M+2$. By inspecting the structure of the Hilbert space, cf. Fig.~\ref{fig:1}a, one can define a unitary transformation of the degenerate eigenstates (\ref{eq:eigenvecs_triangle2}) such that the new states correspond to the scars such as the projected Onsager scar - highlighted in the Fig.~\ref{fig:1}a or the scars appearing in the time-averaged occupations, Fig.~\ref{fig:1}b,c. Specifically, for $M$ even all of these states have the same $|{\cal S}|=M$ while for $M$ odd, there are $\left \lceil M/2 \right \rceil - 1$ zero-energy eigenstates with $|{\cal S}|=M$ and one with $|{\cal S}|=(M+1)/2$, which corresponds to the scar appearing on the diagonal $(0,0) - ((M-1)/2,(M-1)/2)$, cf. Fig.~\ref{fig:1}b. It is apparent that these scar eigenstates are sparse according to Eq.~(\ref{eq:sparsity})
\beq
	\mathscr{S}_{\ket{v_i}} = \frac{\frac{M(M+1)}{2} - O(M)}{\frac{M(M+1)}{2}} \overset{M \rightarrow \infty}{\rightarrow} 1.
	\label{eq:sparsity2}
\eeq


\section{Harmonic approximation and Gaussian integration}
\label{app:gaussian_approx}

A standard assumption in a quench protocol for atoms (or ions) with disorder is to consider an initial state of the form $\rho_{\rm in} = \ket{\psi(0)}\bra{\psi(0)} \otimes \rho_{\rm m}$, where $\rho_{\rm m} = {\rm e}^{-\beta H_{\rm tr}}/{\rm Tr} \left({\rm e}^{-\beta H_{\rm tr}}\right)$ is the mixed state of the motional degrees of freedom corresponding to the trap Hamiltonian $H_{\rm tr}=\sum_k \sum_{\nu=x,y,z} m \omega_\nu^2 \nu_k^2/2$ and $\ket{\psi(0)}\bra{\psi(0)}$ is the pure state of the spins \cite{Gambetta_2020_PRL}. 

Let's next consider a chain of $M$ atoms which are either all in the ground $(\kappa=0)$ or excited (Rydberg) state $(\kappa=1)$. We also assume that each atom experiences the same trapping potential described by $H_{\rm tr}$. This is motivated by the ongoing experimental efforts in trapping the atoms once they are excited in their Rydberg state in optical tweezer setups \cite{Wilson_2019} and we note this assumption has been used in other theory works dealing with the motion of the Rydberg atoms \cite{Gambetta_2020_PRL}. We thus define a classical potential Hamiltonian
\beq
	H_{\rm cl}(\kappa) = H_{\rm tr} + \kappa \delta V,
\eeq
where 
\beqa
	\delta V &=& \sum_{k=1}^{M-1} \frac{C_\alpha}{|{\bf r}_{k+1} - {\bf r}_k|^\alpha} - V_{\rm NN}
	\nonumber \\
	&=& \left. \sum_{k=1}^{M-1} \frac{C_\alpha}{(\lambda^2 \delta x_k^2 + \lambda^2 \delta y_k^2 + (r_0 + \lambda \delta z_k)^2 )^\frac{\alpha}{2}} \right|_{\lambda=1} - V_{\rm NN}
	\nonumber \\
	& \approx & \sum_{k=1}^{M-1} \frac{\alpha C_\alpha}{2 r_0^{\alpha+2}} \left[-2 r_0 \delta z_k - (\delta x_k^2 + \delta y_k^2) + (1+\alpha)\delta z_k^2 \right] + O(\lambda^3)
	\nonumber \\
	&=& V_{\rm NN} \frac{\alpha}{2}\frac{1}{r_0^2} \times \nonumber \\
	&& \left[ -(x_2 - x_1)^2 - (x_3 - x_2)^2 - \ldots -(x_M - x_{M-1})^2 \right. \nonumber \\
	&& \phantom{a}		  -(y_2 - y_1)^2 - (y_3 - y_2)^2 - \ldots -(y_M - y_{M-1})^2	\nonumber \\
	&& \phantom{a}		  +(1+\alpha) \left[ (z_2 - z_1)^2 + (z_3 - z_2)^2 + \ldots +(z_M - z_{M-1})^2 \right] \nonumber \\
	&& \phantom{a}	\left.	- 2 r_0 \left[ (z_2 - z_1) + (z_3 - z_2) + \ldots +(z_M - z_{M-1}) \right]		  
		\right] \nonumber \\
	&=& \frac{1}{2 \beta} {\bf R}^T A_V {\bf R} - \frac{1}{\beta} {\bf B}^T {\bf R} \nonumber \\
	&=& \frac{1}{2 \beta} ({\bf R} - \mu_V)^T A_V ({\bf R} - \mu_V) - \frac{1}{2 \beta} {\bf B}^T A_V^{-1} {\bf B}.
	\label{eq:delta_V_app}
\eeqa
Here $\delta \nu_k = \nu_{k+1} - \nu_k$, $\nu=x,y,z$ and in the third line we have expanded to second order in the small parameter $\lambda$. In the last two lines, ${\bf R} = (x_1,\ldots,x_M,y_1,\ldots,y_M,z_1,\ldots,z_M)^T$, $\mu_V \equiv A_V^{-1} {\bf B}$ and we have introduced $A_V$ and ${\bf B}$, see Eqs.~(\ref{eq:AVxy})-(\ref{eq:Bz}) below for definitions. In words, for $\kappa=1$ the motion of the atoms is a result of the effect of the trapping potential combined with the mutual interactions between the Rydberg atoms [where only nearest neighbour interactions are considered, in agreement with the assumptions of the validity of $H_{\rm eff}$, Eq.~(\ref{eq:Heff})]. 

We define the following probability distributions of the atomic positions
\begin{subequations}
	\begin{align}
	p({\bf R}) & \equiv  \frac{{\rm e}^{-\beta H_{\rm cl}(\kappa=0)}}{{\rm Tr} \left( {\rm e}^{-\beta H_{\rm cl}(\kappa=0)} \right)}
	\propto  {\rm e}^{-\frac{1}{2} {\bf R}^T A_{\rm tr} {\bf R}}
	\label{eq:p}	
	\\	
	p_{\rm int}({\bf R}) & \equiv  \frac{{\rm e}^{-\beta H_{\rm cl}(\kappa=1)}}{{\rm Tr} \left( {\rm e}^{-\beta H_{\rm cl}(\kappa=1)} \right)}
	\propto {\rm e}^{-\frac{1}{2} {\bf R}^T A {\bf R} + {\bf B}^T {\bf R}}.
	\label{eq:p_int}	
	\end{align}
\end{subequations}
In Eqs.~(\ref{eq:p}), (\ref{eq:p_int}), $A=\oplus_{\nu=x,y,z} A^{(\nu)}$, ${\bf B} = \oplus_{\nu=x,y,z} {\bf B}^{(\nu)}$ are block diagonal matrix and vector respectively such that $A^{(\nu)} = A_{\rm tr}^{(\nu)} + A_V^{(\nu)}$ with components
\begin{subequations}
	\begin{align}
		A^{(\nu)}_{\rm tr} &= \beta m \omega_\nu^2 \, {\mathds 1}_{M \times M} = \sigma_\nu^{-2} \, {\mathds 1}_{M \times M}  \\		
		A_V^{(x)} = A_V^{(y)} & = -V_{\rm NN}\frac{\alpha \beta}{r_0^2} {\cal A}_V \label{eq:AVxy} \\
		A_V^{(z)} & = V_{\rm NN}\frac{\alpha \beta}{r_0^2} (1+\alpha) {\cal A}_V \label{eq:AVz} \\
		{\bf B}^{(x)} = {\bf B}^{(y)} & = {\bf 0}_M^T \label{eq:Bxy}\\
		{\bf B}^{(z)} &= V_{\rm NN}\frac{\alpha \beta}{r_0} (-1,0,\ldots,0,1)_M^T. \label{eq:Bz}
	\end{align}
\end{subequations}
In Eqs.~(\ref{eq:Bxy}), (\ref{eq:Bz}), the vectors are of length $M$ and ${\bf 0}_M$ is a zero vector. In Eqs.~(\ref{eq:AVxy}),(\ref{eq:AVz}) ${\cal A}_V$ is a $M \times M$ tridiagonal matrix
\beq
	{\cal A}_V = 
	\begin{pmatrix}
		1 & -1 & & & \\
		-1 & 2 & \ddots & & \\
		 & \ddots & \ddots & \ddots & \\
		 & & \ddots & 2 & -1 \\
		 & & & -1 & 1
	\end{pmatrix}_{M \times M}.
\eeq
	
\begin{figure}[h]
	\centering
	\includegraphics[width=0.8\textwidth]{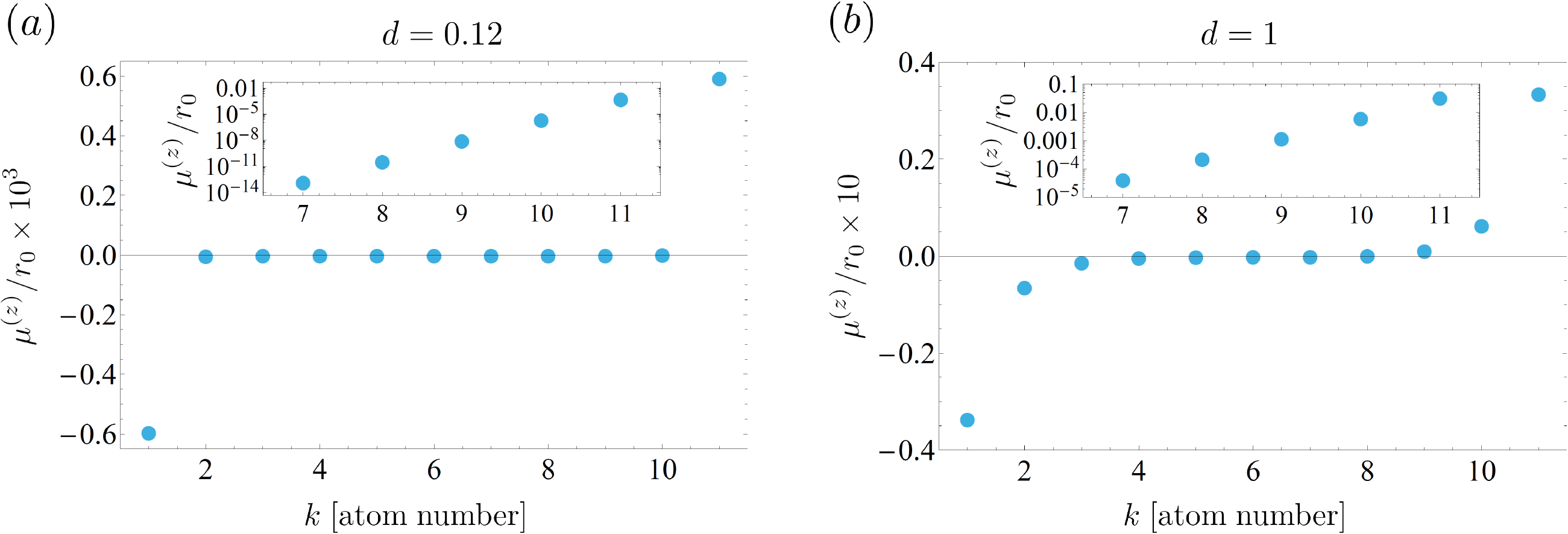}
	\caption{
	Equilibrium atomic positions $\mu^{(z)}$ along the $z$-axis for $M=11$ and (a) $d=0.12$ and (b) $d=1$. The insets show the positions of the atoms of a half chain $k > \left \lceil M/2 \right \rceil$. Motivated by the experimental values from \cite{Marcuzzi_2017_PRL}, $T = 50 \, \mu{\rm K}$, $\omega_z = 2\pi \times 91.4 \, {\rm kHz}$, $m=87 m_u$, where $m_u$ is the atomic mass unit, we set $s_0 = 0.03$.
	}
	\label{fig:mu_atom}
\end{figure}

	The probability distributions (\ref{eq:p}), (\ref{eq:p_int}) can be written, including the normalization factors, as
\begin{subequations}
	\begin{align}
	p({\bf R}) & = \frac{1}{\sqrt{(2\pi)^{3M} |A_{\rm tr}^{-1}|}} e^{-\frac{1}{2}{\bf R}^T A_{\rm tr} {\bf R}}.
	\label{eq:p_norm}	
	\\	
	p_{\rm int}({\bf R}) & = \frac{1}{\sqrt{(2\pi)^{3M} |A^{-1}|}} e^{-\frac{1}{2}({\bf R}-\mu)^T A ({\bf R}-\mu)},
	\label{eq:p_int_norm}	
	\end{align}
\end{subequations}
	where $\mu = (\mu_1,\ldots,\mu_{3M})$ with $\mu_j \equiv \int {\rm d}{\bf R} \, p_{\rm int}({\bf R}) R_j$ and ${\rm d}{\bf R} = \prod_{\nu=x,y,z}\prod_{k=1}^{M} {\rm d}\nu_k$. In particular, comparing (\ref{eq:p_int}) with (\ref{eq:p_int_norm}) and using the fact that $A=A^T$, we get $\mu = A^{-1} B$, where zero values are implicitly assumed for the singular part of $A^{-1}$ corresponding to the $x,y$ blocks. As a result, only $\mu^{(z)} \neq {\bf 0}_M$ and an example of the equilibrium atomic positions in the presence of (repulsive) interactions is shown in Fig.~\ref{fig:mu_atom}. It is apparent that in an open chain considered here, the presence of interactions is mostly affecting the outermost atoms. 
	
With the above definitions at hand, we are in position to evaluate the expectation values of the disorder energies. 
Using the relation
\begin{equation}
	\int {\rm d}{\bf R} \, \frac{1}{\sqrt{(2\pi)^{3M} |C^{-1}|}} \, e^{-\frac{1}{2}({\bf R}-\mu)^T C ({\bf R}-\mu)} ({\bf R}-\eta)^T D ({\bf R}-\eta)
	= {\rm Tr}(C^{-1}D)+ (\mu-\eta)^T D (\mu-\eta),
\end{equation}
where $C,D$ are $3M \times 3M$ matrices, we will specifically evaluate the expectation values with the probabilities (\ref{eq:p_norm}), (\ref{eq:p_int_norm}). The results read
\begin{subequations}
	\begin{align}
		\braket{\delta V} &\equiv \int {\rm d}{\bf R} \, p({\bf R}) \delta V = \frac{1}{2\beta}{\rm Tr}(A_{\rm tr}^{-1} A_{V})
		=   (M-1) \alpha V_{\rm NN} \left[ (1+\alpha) s_z^2 - (s_x^2 + s_y^2) \right] 
		\label{eq:exp_delta_V}\\
		\braket{\delta V}_{\rm int} &\equiv \int {\rm d}{\bf R} \, p_{\rm int}({\bf R}) \delta V = \frac{1}{2\beta}\left({\rm Tr}(A^{-1} A_{ V}) + (\mu - 2 \mu_V) A_{V} \mu \right).
		\label{eq:exp_delta_V_int}
	\end{align}
\end{subequations}
In Eq.~(\ref{eq:exp_delta_V}), $s_\nu = \sigma_\nu/r_0$, where $\sigma_\nu = \sqrt{1/(\beta m \omega_\nu^2)}$. Using the parametrization of the disorder ${\bf s} = (\epsilon,1,1) d s_0$ then leads to the Eq.~(\ref{eq:Vstep}). Since $A,A_V$ are symmetric tridiagonal matrices, they can be diagonalized analytically \cite{Kouachi_2006, Yueh_2005, Kilic_2008, Kulkarni_1999, Mallik_2001, Banchi_2013} and thus in principle evaluate also the Eq.~(\ref{eq:exp_delta_V_int}), yielding nevertheless rather cumbersome expressions. For this reason we evaluate Eq.~(\ref{eq:exp_delta_V_int}) numerically.

\section{Hilbert space energy landscape}

\begin{figure}[t!]
	\centering
	\includegraphics[width=0.65\textwidth]{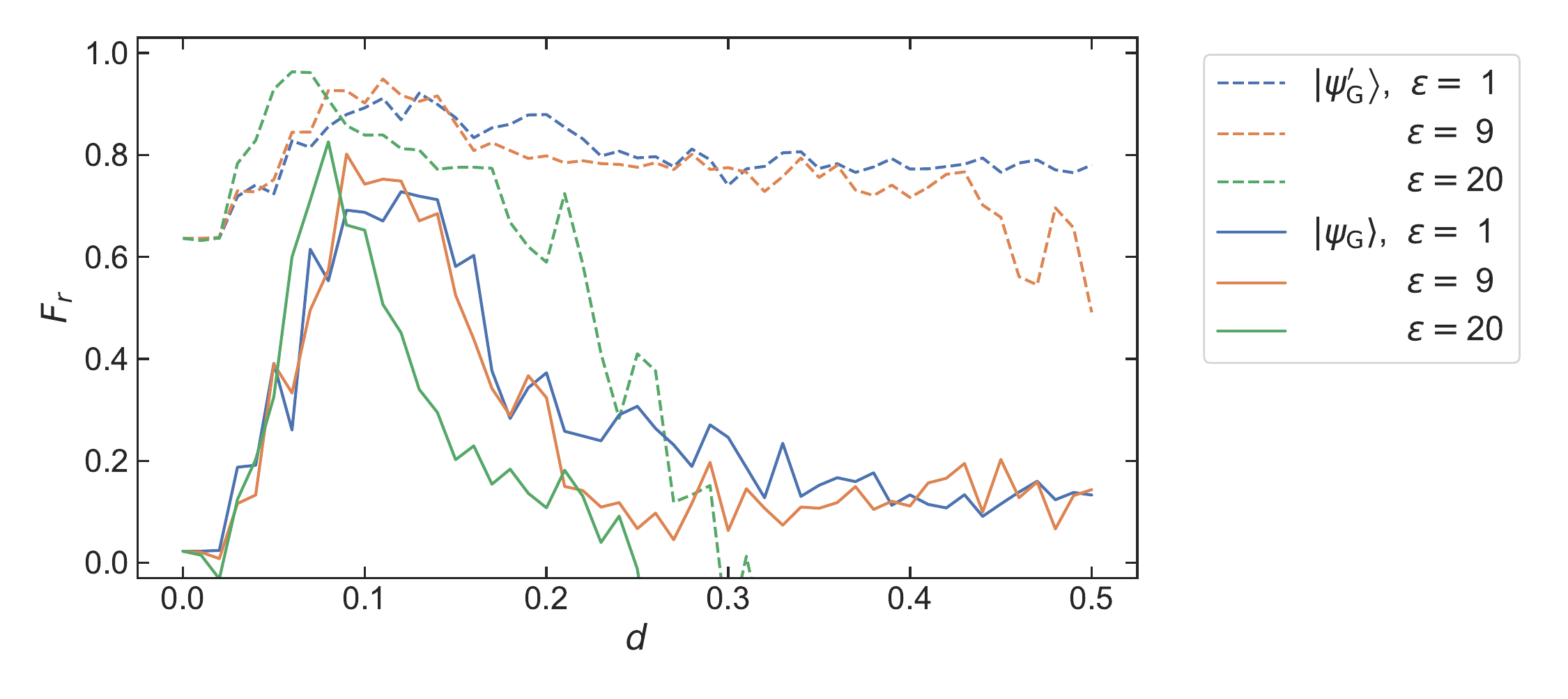}
	\caption{The rescaled fidelity $F_r$ vs. the disorder strength $d$ for various anisotropies of the traps $\epsilon$ (blue, orange and green for $\epsilon=1,9$ and 20) and initial states $\ket{\psi_G}$ (solid lines) and $\ket{\psi_G'} \equiv \ket{\psi_{\bar{\bf x}_0 = (0,4)}^{{\bf p}=(\pi/2,-\pi/2),w=2}}$ (dashed lines) [cf. Figs.~\ref{fig:1}b,c]. The results are for $M=11$ and 10 realizations of the initial conditions.}	
	\label{fig:S_Fidelity}
\end{figure}

In Fig.~\ref{fig:2}f of the main text, we have evaluated the fidelity $F_r$ of the QMBS obtained with the Hamiltonian (\ref{eq:H}) compared to the ones obtained with $H_0$, Eq.~(\ref{eq:H0}), corresponding to the (truncated) square lattice where all sites (basis states) have the same energies. We note that the addition of the potential term $H_{\rm pot}$, Eq.~(\ref{eq:Hpot}), creates a characteristic staircase potential with energy difference of $V_{\rm NNN}$ between the adjacent diagonals $(0,m)-(m,0), m=0..M-2$ (there is no energy difference, in the absence of disorder, between the $(0,M-1)-(M-1,0)$ and $(0,M-2)-(M-2,0)$ diagonals corresponding to blocks of a single and two consecutive up spins respectively). 
We next define the total number of nearest and next-to-nearest neighbours $n_{\rm NN}=\sum_{k} n_k n_{k+1}$, $n_{\rm NNN}=\sum_k n_k n_{k+2}$ and $\braket{\delta V(n_{\rm NN})}$ to be the expectation value of $\delta V_{\bar{\bf x}}$, Eq.~(\ref{eq:delta_V}), corresponding to basis state $\ket{\bar{\bf x}}$ containing $n_{\rm NN}$ nearest neighbours and evaluated using $p(\delta {\bf r}_k)$.
To characterize the effect of the disorder, we define
\beqa
	V_{\rm step}(n_{\rm NN}) &=& V_{\rm NNN} + \braket{\delta V(n_{\rm NN})} - \braket{\delta V(n_{\rm NN}-1)} \nonumber \\
			&=& V_{\rm NNN} + V_{\rm NN} \alpha (\alpha-\epsilon^2) (d s_0)^2,
	\label{eq:Vstep}
\eeqa
where we have used the result (\ref{eq:exp_delta_V}). Consequently, if $\epsilon > \sqrt{\alpha}$, $V_{\rm step}=0$ corresponds to an \emph{on-average} flattening of the potential by disorder, which occurs for $s_0 d^* = \sqrt{V_{\rm NNN}/(V_{\rm NN} \alpha (\epsilon^2-\alpha))} = \sqrt{1/(2^\alpha \alpha (\epsilon^2-\alpha))}$. 
It is thus tempting to assume that such an average flattening might be related to the enhancement of the scar behaviour as quantified by $F_r$. Here we argue that this is not the case based on further numerical evidence and analysis of the disorder properties.

In Fig.~\ref{fig:S_Fidelity} we present the results of $F_r$ for initial states $\ket{\psi_G}$ used in the main text and $\ket{\psi_G'}$, cf. Fig.~\ref{fig:1}c and the caption of the Fig.~\ref{fig:S_Fidelity} for the definition, and various values of the trap anisotropy parameter $\epsilon$. It is obvious from Fig.~\ref{fig:S_Fidelity} that the enhancement of $F_r$ occurs for all values of $\epsilon$, including the isotropic traps $\epsilon=1$ for which there is no on-average flattening according to (\ref{eq:Vstep}) (for $\alpha=6$ used here). We also note the higher $F_r$ for the initial state $\ket{\psi_G'}$ (except for $\epsilon=20$ and $d \gtrsim 0.25$). This is in line with the argument that $\ket{\psi_G'}$ effectively approaches the Onsager scar state $\ket{\psi} \propto \sum_k (-1)^k \sigma^+_k \ket{\da \ldots \da}$ (projected on the $N_{\rm cl}=1$ sector), cf. Fig.~\ref{fig:1}a, which is an exact eigenstate of $H$, independent of disorder and hence for which $F_r = F = 1$.

We now analyse the properties of the disorder appearing in Eq.~(\ref{eq:Vstep}), in particular the expectation values of the first and second moments of the interaction energies corresponding to the mean value and the width of the respective distributions. From Eq.~(\ref{eq:delta_V}) we have
\be
	\delta V(n_{\rm NN}) - \delta V(n_{\rm NN} - 1) = \left.  V_{\rm NN} \, \alpha \lqq - \tilde{\delta}_{z,k} - \ha \lt \tilde{\delta}_{x,k}^2 + \tilde{\delta}_{y,k}^2 - (1+\alpha) \tilde{\delta}_{z,k}^2  \rt   \rqq \right|_{k = \bar{x} + 1 + n_{\rm NN}}.
\ee
Since the distribution $p(\delta \mathbf{r}_k)$ over which we average does not explicitly depend on $k$ (i.e., the disorder is translationally invariant), we drop the index $k$ for brevity. We get for the expectation values
\be
	\av{\tilde{\delta}_\nu} = 0  \comma \av{\tilde{\delta}_y^2} = \av{\tilde{\delta}_z^2} = 2 (s_0 d)^2 \mand   \av{\tilde{\delta}_x^2} = 2 \epsilon^2 (s_0 d)^2.
\ee
We now introduce the dimensionless shorthands 
\beq
	\chi = \frac{1}{2 (s_0 d)^{2}}
	\label{eq:chi_def}
\eeq
and
\be
	\delta v = \frac{\delta V(n_{\rm NN}) - \delta V(n_{\rm NN} - 1)}{V_{\rm NN}},
\ee
so that
\be
	\av{\delta v} = -\ha \alpha \lqq \av{\tildel_x^2} + \av{\tildel_y^2} - (1+\alpha) \av{\tildel_z^2}  \rqq =  -\ha \frac{\alpha}{\chi} \lqq \epsilon^2 + 1 - (1+\alpha)  \rqq = \frac{\alpha}{2\chi} \lt   \alpha - \epsilon^2 \rt.
\ee
The threshold value $s_0 d^*$ introduced after Eq.~(\ref{eq:Vstep}) corresponds to $V_{\rm NN} \av{\delta v} = -V_{\rm NNN} = -2^{-\alpha} V_{\rm NN} $, i.e., $\av{\delta v} = -2^{-\alpha}$, fixing in turn
\be
	\chi = 2^{\alpha - 1} \alpha \lt   \epsilon^2 - \alpha  \rt.
	\label{eq:thresh1}
\ee
In words, $V_{\rm NN} \av{\delta v}$ is centred around $-V_{\rm NNN}$. However, is it \emph{peaked} around this value? In order to better understand this, we compute the variance of $V_{\rm NN} \delta v$ and compare it to $V_{\rm NNN}^2$. We start with
\be
	\frac{\delta v^2}{\alpha^2} = \tildel_z^2 + \tildel_z \lt \tildel_x^2 + \tildel_y^2 - (1+\alpha) \tildel_z^2  \rt + \frac{1}{4} \lt \tildel_x^2 + \tildel_y^2 - (1+\alpha) \tildel_z^2  \rt^2
\ee
and recall that for zero-mean, independent Gaussian variables $\av{\tildel_\nu^4} = 3 \av{\tildel_\nu^2}^2$. Hence, we find
\be
	\av{\frac{\delta v^2}{\alpha^2}} = \av{\tildel_z^2}  + \frac{1}{4 } \lqq 2\av{\tildel_x^2}^2 + 2\av{\tildel_y^2}^2 + 2(1+\alpha)^2 \av{\tildel_z^2}^2 +  \lt \av{\tildel_x^2} + \av{\tildel_y^2} - (1+\alpha) \av{\tildel_z^2}  \rt^2 \rqq.
\ee
Note that the final addend in the brackets yields, once the multiplicative constants are accounted for, the squared average of $\delta v / \alpha$ and therefore
\be
	\av{\frac{\delta v^2}{\alpha^2}} - \av{\frac{\delta v}{\alpha}}^2 = \av{\tildel_z^2}  + \frac{1}{4 } \lqq 2\av{\tildel_x^2}^2 + 2\av{\tildel_y^2}^2 + 2(1+\alpha)^2 \av{\tildel_z^2}^2  \rqq = \frac{1}{\chi} + \frac{1}{2 \chi^2} \lqq \epsilon^4 + 1 + (1+\alpha)^2   \rqq.
	\label{eq:var}
\ee
To quantify the width of the distribution of energies $V_{\rm NN} \delta v$ with respect to its centre $-V_{\rm NNN}$ we define the ratio
\be 
	\eta^2 
	= \frac{\av{(V_{\rm NN} \delta v)^2} - \av{V_{\rm NN}\delta v}^2}{V_{\rm NNN}^2}
	= 2^{2\alpha} \left( \av{\delta v^2} - \av{\delta v}^2 \right)
	=\frac{2^{2\alpha}}{2\chi^2} \alpha^2 \lqq 2\chi + \epsilon^4 + 1 + (1+\alpha)^2   \rqq,
\ee
where in the last equality we have substituted from Eq.~(\ref{eq:var}). $\eta$ thus represents the standard deviation of the energy distribution such that we can write, with a slight abuse of notation,
\be
	\delta V(n_{\rm NN}) - \delta V(n_{\rm NN} - 1) \approx - V_{\rm NNN} \pm \eta V_{\rm NNN}.
	\label{eq:V_width}
\ee
We will now provide a numerical example. Here, it is worth noting that according to the Eq.~(\ref{eq:Vstep}), the isotropic disorder $\epsilon=1$ does not fulfill the necessary condition $\epsilon^2 > \alpha$ and thus does not lead the the on-average flattening. Taking $\epsilon=9$, we find $d^* \approx 0.2$ which yields $\eta \approx 3.5$ (we recall we use $s_0 = 0.03$ motivated by \cite{Marcuzzi_2017_PRL}). It is thus apparent from (\ref{eq:V_width}), that for the values of the disorder for which one gets the on-average flattening $V_{\rm step} \approx 0$, the width of the energy distribution has already become much broader than its mean value making thus the flattening argument effectively irrelevant. \\
~\\

A different calculation could be set up to assess for what value of the disorder strength the distribution becomes sufficiently broad to render the bias $- V_{\rm NNN}$ effectively irrelevant. One way to look for such a threshold is to ask when the centre of the shifted distribution $\av{V_{\rm NN} \delta v} + V_{\rm NNN}$ is of the same order of the standard deviation, which can also be written as
\be 
	\lqq  \av{\delta v} + \mu  \rqq^2 = \av{\delta v^2} - \av{\delta v}^2.
\ee
This corresponds to
\be
	\lqq \frac{\alpha}{2\chi}(\alpha - \epsilon^2)  + 2^{-\alpha}    \rqq^2 = \frac{\alpha^2}{2\chi^2} \lqq  2\chi + \epsilon^4 + 1 + (1+\alpha)^2 \rqq .
\ee
Multiplying both sides by $(2\chi/\alpha)^2$ we get a quadratic polynomial in $\chi$
\be
	a_2 \chi^2 + a_1 \chi + a_0 = 0
\ee
with
\begin{subequations}
\begin{align}
	a_2 & = \frac{2^{2-2\alpha}}{\alpha^2} > 0 \\
	a_1 & = \frac{2^{2-\alpha}}{\alpha} \lt\alpha - \epsilon^2 \rt - 4 \\
	a_0 & = \lt\alpha - \epsilon^2 \rt^2 - 2\epsilon^4 - 2 - 2(1+\alpha)^2 < 0.
\end{align}
\end{subequations}
By Descartes' rule of signs, there are always a positive and a negative solution. By its definition, $\chi$ must be positive, so we can discard the negative one. Hence, the threshold value we are looking for can be written as a function of $\alpha$ and $\epsilon$ in the combination
\be
	\chi_{\rm th}(\alpha, \epsilon) = \frac{1}{2a_2} \lqq -a_1 + \sqrt{a_1^2 - 4a_2 a_0}  \rqq.
\ee
Applying this to the case $\epsilon=9$ and using the definition of $\chi$ Eq.~(\ref{eq:chi_def}) we find that the width of the distribution becomes comparable to $V_{\rm step}$ for $d \approx 0.06$.

\section{Numerical treatment of the atom motion}

We model the atomic motion as that of a classical point particle in a harmonic trap with coordinates $\nu(t) = C_{\nu} \cos(\omega_\nu t + \phi_{\nu})$, 
where 
$C_{\nu} = \sqrt{\nu(0)^2 + (q_{\nu}(0)/m)^2 / \omega_\nu^2)}$ 
and
$\phi_{\nu} = \arccos\left(\nu(0) / C_{\nu}\right)$
which are fully specified by the initial position $\nu(0)$ and velocity $v_{\nu}(0) = q_{\nu}(0)/m$ for each of the direction $\nu=x,y,z$. These are solutions of the equations of motion corresponding to the classical single-particle Hamiltonian
\beq
	h_{\nu}(\nu, v_\nu) = q_\nu^2/2m + 1/2 m \omega_\nu^2 \nu^2. 
\eeq
In the numerical procedure, we draw the initial positions for each direction $\nu$ and velocity $v_\nu$ from the corresponding Gaussian (Boltzmann) probability distribution
\beq
	p(\nu,v_\nu) = \frac{{\rm e}^{-\beta h_\nu}}{\cal Z}
	= p(\nu) p(v_\nu)
	  = \frac{{\rm e}^{-\frac{\nu^2}{2 \sigma_\nu^2}}}{\sqrt{2\pi}\sigma_\nu} \frac{{\rm e}^{-\frac{v_\nu^2}{2 \sigma_{v,\nu}^2}}}{\sqrt{2\pi}\sigma_{v,\nu}},
	  \label{eq:pnu}
\eeq
where ${\cal Z} = {\rm Tr} \left( {\rm e}^{-\beta h_\nu} \right)$, $\sigma_\nu^2 = 1/(\beta m \omega_\nu^2)$ and $\sigma_{v,\nu} = \omega_\nu \sigma_\nu$. For later convenience we also introduce the corresponding functions of the momenta $q_\nu = m v_\nu$ rather than the velocities
\beqa
	H_{\nu}(\nu, q_\nu) &=& h_\nu \left(\nu, \frac{q_\nu}{m} \right) \\
	P(\nu, q_\nu) &=& \frac{{\rm e}^{-\beta H_\nu}}{Z} = \frac{1}{m} p \left(\nu, \frac{q_\nu}{m} \right),
\eeqa
where $Z = m {\cal Z}$.

Due to the nature of the probability distributions there is a possibility of a rare event when two atoms come arbitrarily close to each other resulting in the distribution of interaction energies with no defined moments \cite{Ostmann_2019_QSciTech}. It is thus instructive to investigate what is the probability of such an event if one  imposes a threshold on the initial conditions, namely
\beq
	h_\nu \leq E,
	\label{eq:constraint}
\eeq
where $E$ is a cutoff energy. We define the \emph{acceptance} probability as
\be
	P_{\rm acc}^{(\nu)} = \int_{H_\nu \leq E} \rmd \nu\, \rmd q_\nu \,\, \frac{\rme{- \beta H_\nu }}{Z} = \frac{\int_{H_\nu \leq E} \rmd \nu\, \rmd q_\nu \,\, \rme{- \beta H_\nu }}{\int \rmd \nu\, \rmd q_\nu \,\, \rme{- \beta H_\nu }}  .
	\label{eq:Pacc}
\ee

We now apply the standard canonical transformation to action-angle coordinates $(Q_\nu, K_\nu)$ for the harmonic oscillator
\be
\begin{split}
	\nu & = \sqrt{\frac{2 K_\nu}{m\omega_\nu}} \sin{Q_\nu} \\
	q_\nu & = \sqrt{2 m \omega_\nu K_\nu} \cos{Q_\nu}
\end{split}
\ee
Because the transformation is canonical, the Jacobian of the change of variables corresponds to the Poisson brackets:
\beq
	{\cal J} = \left| \begin{matrix} \frac{\partial \nu}{\partial Q_\nu} & \frac{\partial \nu}{\partial K_\nu}  \\[2mm] \frac{\partial q_\nu}{\partial Q_\nu} & \frac{\partial q_\nu}{\partial K_\nu} \end{matrix} \right| 
	= \left| \frac{\partial \nu}{\partial Q_\nu}\, \frac{\partial q_\nu}{\partial K_\nu} - \frac{\partial \nu}{\partial K_\nu} \, \frac{\partial q_\nu}{\partial Q_\nu} \right| 
	= \left| \{ \nu, q_\nu \}_{\rm Poisson} \right| = 1. 
\eeq
Additionally,
\be
	K_\nu = \frac{H_\nu}{\omega_\nu}
\ee
so that (\ref{eq:Pacc}) becomes
\be
	P_{\rm acc}^{(\nu)} = \frac{\int_{K_\nu \leq E / \omega_\nu} \rmd Q_\nu\, \rmd K_\nu \,\, \rme{- \beta \omega_\nu  K_\nu }}{\int \rmd Q_\nu\, \rmd K_\nu \,\, \rme{- \beta \omega_\nu K_\nu }} .
	\label{eq:Pacc2}
\ee
Since neither the constraint nor the integrand depend on $Q_\nu$, (\ref{eq:Pacc2}) evaluates to
\be
	P_{\rm acc}^{(\nu)} = \frac{\lt 1/ (\beta\omega_\nu) \rt \lt 1 - \rme{-\beta E}  \rt }{\lt 1/ (\beta\omega_\nu) \rt} = 1 - \rme{-\beta E}.
	\label{eq:Pacc3}
\ee

Alternatively, this result can be obtained by direct evaluation using the probability distribution Eq.~(\ref{eq:pnu}) by noting that the constraint (\ref{eq:constraint}) is nothing but a definition of the disk
\beq
	\tilde{v}_\nu^2 + \tilde{\nu}^2 \leq 1
	\label{eq:disk}
\eeq
upon the obvious change of variables. In this case the acceptance probability reads
\beq
	P^{(\nu)}_{\rm acc} \equiv \int_{\cal D} {\rm d}\nu \, {\rm d}v_\nu \, p(\nu, v_\nu),
\eeq
where the integration is performed over the disk defined by (\ref{eq:disk}). Transforming $\tilde{\nu}, \tilde{v}_\nu$ to polar coordinates $\tilde{\nu} = \tilde{r} \cos \varphi, \; \tilde{v}_\nu = \tilde{r} \sin \varphi$ and integrating first over the angles yields an expression of the form
\beq
	P^{(\nu)}_{\rm acc} \propto 2 \pi \int_0^1 {\rm d}\tilde{r} \tilde{r} {\rm e}^{-\frac{1}{2}(a+b)\tilde{r}^2} I_0\left(\frac{1}{2}(a-b)\tilde{r}^2\right),
\eeq
where $I_0$ is the modified Bessel function and $a=E/(m \omega_\nu^2 \sigma_\nu^2)$, $b=E/(m \sigma_{v,\nu}^2)$. It follows from $\sigma_{v,\nu} = \omega_\nu \sigma_\nu$ that $a=b$ and consequently $I_0\left(\frac{1}{2}(a-b)\tilde{r}^2\right)=1$ with the result (\ref{eq:Pacc3}).

To evaluate numerically the effect of the cutoff, we parametrize the cutoff energy as $E=\frac{1}{2} m \omega_\nu^2 (r_0/f)^2$, such that $f$ describes the fraction of $r_0$ which determines the maximum allowed distance of an atom from the trap center and $s_\nu$ are the disorder strenghts $(s_x,s_y,s_z) = (\epsilon,1,1) d s_0$ using the notation of the main text. Defining the \emph{rejection} probability 
\beq
	P^{(\nu)}_{\rm rej} = 1 - P^{(\nu)}_{\rm acc} = {\rm e}^{-\frac{E}{m \omega_\nu^2 \sigma_\nu^2}} = {\rm e}^{-\frac{1}{2 (f s_\nu)^2}}.
	\label{eq:Prej}
\eeq
it follows that for $f \rightarrow 0$ (arbitrarily high cutoff energy) and $s_\nu \rightarrow 0$ (no disorder), the rejection probability vanishes as it should (and similarly $P_{\rm rej} \rightarrow 1$ for $f,s_\nu \rightarrow \infty$). 

Finally, we note that only the displacements along the chain axis ($z$-axis) lead to the divergences of the interaction energy when the positions of the two atoms coincide. We thus evaluate (\ref{eq:Prej}) for the largest amount of disorder conisdered $s_z = s_0 d$ for $d=1$ and taking $f=2$, i.e. allowing each atom to be at most the distance $r_0/2$ away from the trap center, which yields $P^{(z)}_{\rm rej} \approx 4.8 \cdot 10^{-61}$. We thus conclude that for the number of realizations $O(100)$ and the parameters considered in this work, the cutoff condition (\ref{eq:constraint}) can be safely neglected.

\subsection{Time evolution of the entanglement entropy}

\begin{figure}[h]
	\centering
	\includegraphics[width=0.65\textwidth]{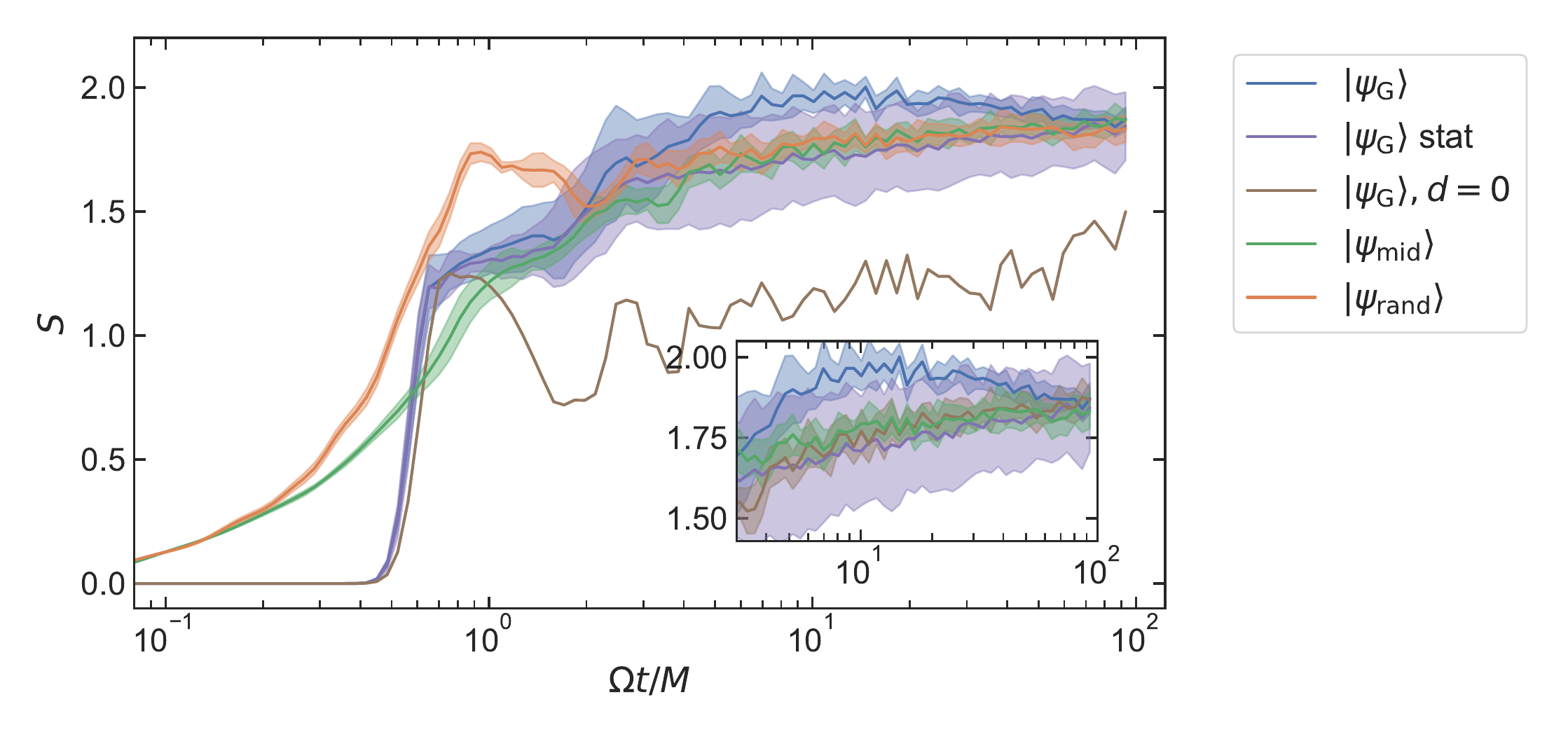}	
	\caption{					
		Evolution of the half-chain entanglement entropy for $M=41$ for $\ket{\psi_G}, \ket{\psi_{\rm mid}}$ and $\ket{\psi_{\rm rand}}$ (solid blue, green and orange lines) defined analogously to the states used in Fig.~\ref{fig:3}. The solid violet (brown) line corresponds to a quench from $\ket{\psi_G}$ with static positional (zero) disorder.
Parameters used: $s_0 = 0.03, \epsilon=9, d=0.12$ and 10 realizations of the disorder (initial conditions). The shaded areas indicate the standard deviation of $S$.		
			}
	\label{fig:S_EE_growth}
\end{figure}

As described in the main text, to extract the (second R\'{e}nyi) entanglement entropy we solve numerically the Schr\"{o}dinger equation with the Hamiltonian Eq.~(\ref{eq:H}), which is explicitly time dependent. 
In Fig.~\ref{fig:S_EE_growth} we show extended data with the same parameters as in Fig.~\ref{fig:3}a but for a larger system size $M=41$. For $\ket{\psi_{\rm mid}}$ and $\ket{\psi_{\rm rand}}$ we observe a slow log-like late-time growth for $\Omega t/M \gtrsim 5$. The projected Gaussian state $\ket{\psi_G}$ on the other hand depicts a faster rise up to a saturation around $\Omega t/M \approx 8$ followed by a decrease for $\Omega t/M > 11$ (cf. also the inset). 


To understand the origin of the decrease of $S$ for $\ket{\psi_G}$, we show the time evolution of $S(t)$ for $\ket{\psi_G}$ with no disorder (solid brown line) and with a \emph{static}, i.e. quenched positional disorder, where the coordinates of each atom are drawn from the distribution $p(\delta {\bf r}_k)$, $k=0,\ldots,M-1$ (solid violet line). In the static disorder case, after the initially similar dynamics, we see a clear departure around $\Omega t/M \approx 3$ followed by a growth which is considerably slower than when accounting for the motion. On the one hand, this highlights the importance of taking the atomic motion into account to faithfully describe a realistic experimental setting. On the other hand, it also shows that the $r$-statistics evaluated with the static positional disorder serves only as an indicator of the integrability properties of the Hamiltonian $H$ when it is driven at the atomic motion frequencies $\omega_\nu$.

\section{Hilbert space structure}
\label{app:Hilbert_space}


\begin{figure}[h]
	\centering
	\includegraphics[width=0.75\textwidth]{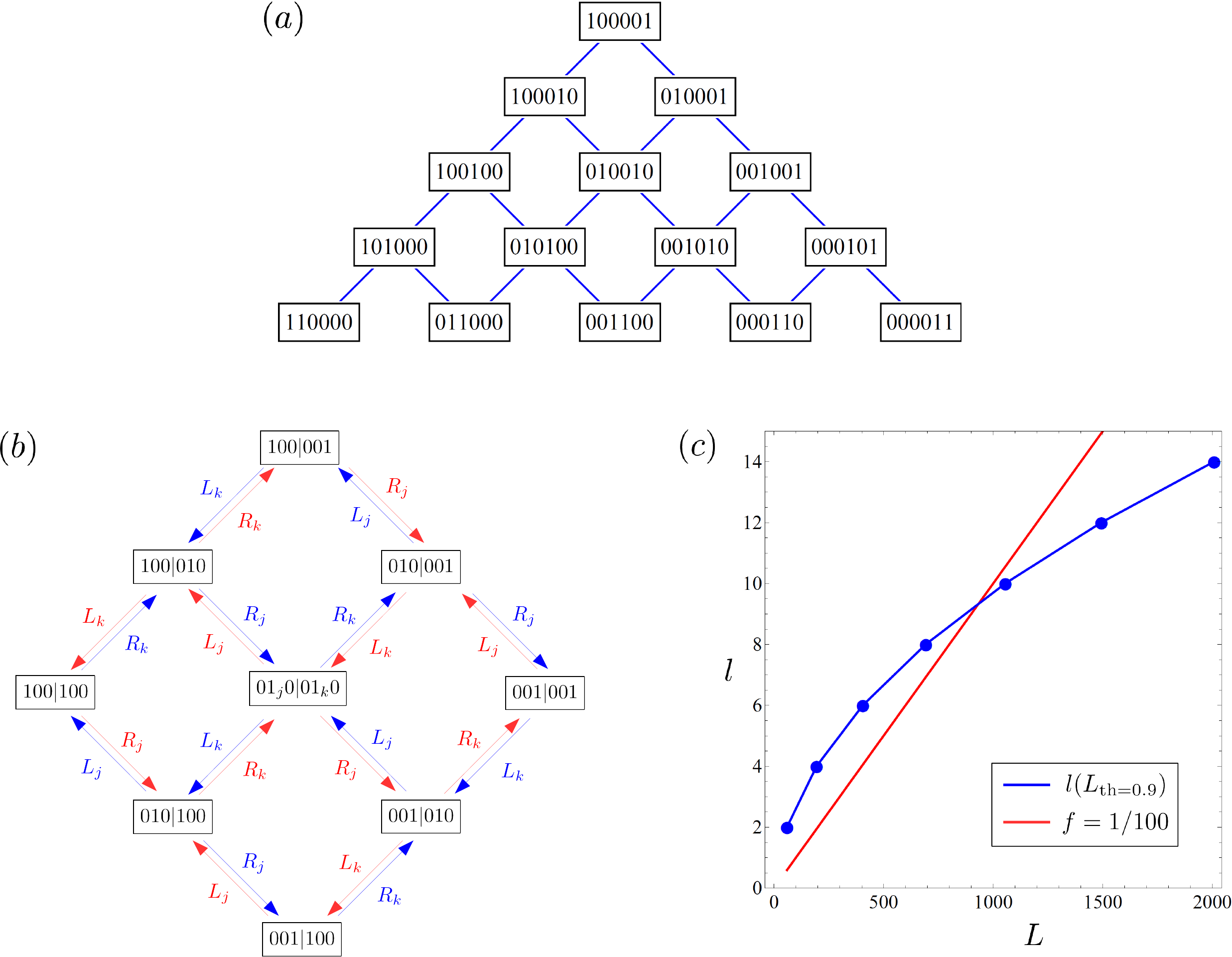}
	\caption{
	{\bf (a)} The structure of the Hilbert space of the XX model, Eq.~(\ref{eq:HXX}), for $l=2$ and $L=6$.
	{\bf (b)} Counting of the loops emanating from the basis state containing a pair $1_j, 1_k$ of up-spins.
	{\bf (c)} Threshold value of $L$ (blue data points) for which $V_{\rm bulk}/V_{\rm boundary} > 0.9$, cf. Eqs.~(\ref{eq:V}). The solid red line indicates the curve of constant filling fraction $f=l/L=1/100$ for comparison.
	}
	\label{fig:Hilbert}
\end{figure}

The spin flip term $\sum_k \sigma^x_k P_{\braket{k}}$ of the effective Hamiltonian Eq.~(\ref{eq:Heff}) on the chain of length $M$ maps to the XX Heisenberg spin-1/2 model of length $L=M+1$, $H_{\rm XX} = \sum_{k=1}^{L-1} \mu^x_k \mu^x_{k+1} + \mu^y_k \mu^y_{k+1}$ - cf. Eq.~(\ref{eq:HXX}) - where $\mu^z = \sum_k \mu^z_k$ is a conserved charge. It is interesting to consider the structure of the corresponding Hilbert space for given system size and number of the Heisenberg excitations, which we denote by $l$, $l=\sum_k 1/2(1+\mu^z_k)$. An example for $l=2$ and $L=6$ is shown in Fig.~\ref{fig:Hilbert}a, which is equivalent to Fig.~\ref{fig:1}a. Here, the Hilbert space structure, in the $\{ \ket{0}, \ket{1} \}$ basis, corresponds to a regular graph (a square lattice), up to the boundaries. This holds for arbitrary $l$ in the limit of vanishing filling fraction as stated in the following lemma:
~\\
~\\
\emph{Lemma:}
The graph topology of the adjacency matrix $H_{\rm XX}$, Eq.~(\ref{eq:HXX}), expressed in the $\{\ket{0},\ket{1}\}$ basis for a fixed $l$ and $L \rightarrow \infty$ corresponds to a hypercubic lattice of dimension $l$. \\
~\\
\emph{Proof:}
The dimension of the Hilbert space of each $l$-sector is given by
\beq
	{\rm dim}_l = \binom{L}{l}.
\eeq
It follows from the particle-hole symmetry of $H_{\rm XX}$ that the sectors $l$ and $l'=L+1-l$ are isomorphic. In the following we thus consider $l < \left \lceil \frac{L}{2} \right \rceil$, i.e. any $l$ below half-filling. 
Next, we define the valency (i.e. the vertex degree) of a basis state $\ket{b_i}$ as the number of connections to other basis states, $v_{\ket{b_i}}= \sum_{j \neq i} \braket{b_i | H_{\rm XX} | b_j}$. It follows that for a given $l$, $v_{\ket{b_i}}$ can take values in $\{1,\ldots, v_{\rm max} \}$, where $v_{\rm max}=2l$. The total number of the basis states with the maximum valency is
\beq
	{\rm dim}_{v_{\rm max}} = \binom{L-1-l}{l-1}.
\eeq
Defining the ratio
\beq
	{\frak r}(l,L) = \frac{{\rm dim_{v_{\rm max}}}}{{\rm dim}_l}
\eeq
in the limit $L \rightarrow \infty$ while keeping $l$ constant, we have
\beq
	\lim_{L \to \infty} \left. {\frak r}(l,L) \right|_{l={\rm const.}} = 1,
\eeq
i.e. the basis states of maximum valency occupy most of the Hilbert space (asymptotically all of it), such that it can be represented as a $v_{\rm max}$-regular graph. Specifically, it corresponds to a \emph{hypercube} of dimension $l$. To show this, we shall count the number of minimal-length loops emanating from a vertex of maximum valency. To this end we first note, that the maximum valency state corresponds to the configuration of the form
\beq
	\ldots 1_j \ldots 1 \ldots 1_k \ldots 1 \ldots,
\eeq
where $\ldots$ stand for string of zeros and there is in total $l$ excitations (ones) which are preceded and followed by at least one zero. In other words there is at least one zero separating two 1s and at least one zero at each end of the chain. The action of $H_{\rm XX}$ is nothing but a permutation $01 \leftrightarrow 10$ shifting a given 1 to either left or right. Denoting such left/right shifts acting on the $j$-th excitation as $L_j, R_j$, a minimal-length loop is formed by interlacing the $L,R$ operations on any pair of excitations $(1_j,1_k)$ as shown in Fig.~\ref{fig:Hilbert}b. As a result, we have four loops of minimal-length four for each pair $(1_j,1_k)$, i.e.
\beq
	\# \; {\text {of loops emanating from each max. valency vertex}} = 4\binom{l}{2},
\eeq
which corresponds to the hypercube (specifically, we get four loops for $l=2$ corresponding to a square lattice and 12 loops for $l=3$ corresponding to the cube). \hfill \emph{QED.} \\
~\\
As a consequence, this allows us to define the volume of the bulk and of the boundary of the Hilbert space as the number of maximum-valency basis states and its complement respectively
\begin{subequations}
	\label{eq:V}
	\begin{align}
		V_{\rm bulk} & \equiv {\rm dim}_{v_{\rm max}} \\
		V_{\rm boundary} &= {\rm dim}_l - V_{\rm bulk}.
	\end{align}
\end{subequations}

\begin{figure}[h]
	\centering
	\includegraphics[width=0.75\textwidth]{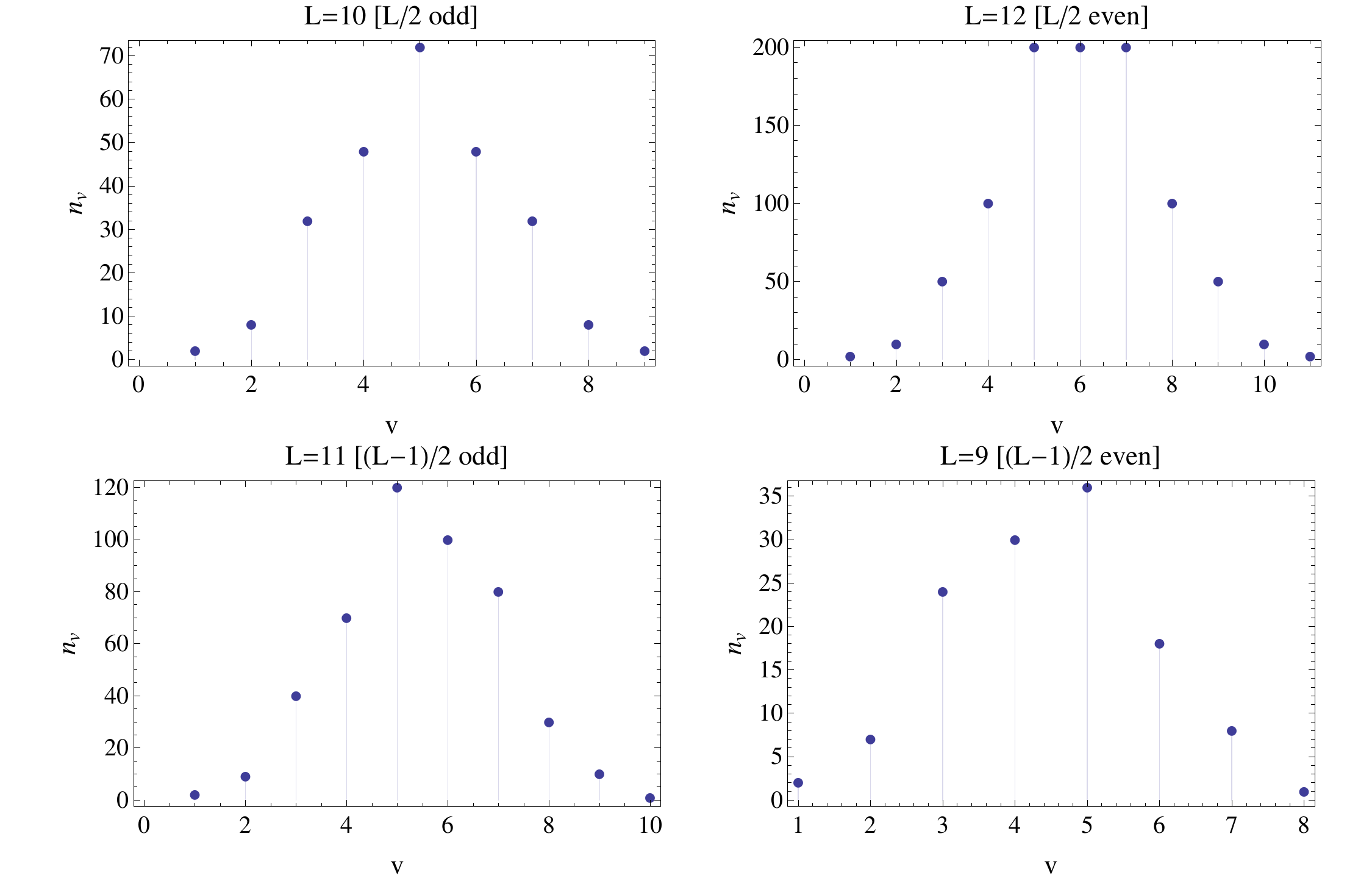}
	\caption{Histograms of the vertex valencies for various $L$ at half-filling.}
	\label{fig:histograms}
\end{figure}

It is interesting to compare the situation of $l={\rm const.}$ to the the constant filling fraction $l/L = {\rm const.}$ instead. In particular, we choose the limiting case of half filling, $\left \lceil l/L \right \rceil = 1/2$. The histogram of the number of vertices of different valencies is shown in Fig.~\ref{fig:histograms}. At half filling, ${\rm dim}_{v_{\rm max}}=1$ (2) for $L$ odd (even) respectively and thus $\lim_{L \to \infty} {\mathfrak r}(\left\lceil \frac{L}{2} \right\rceil,L) = 0$ implying \emph{no volume} in the sense of the definition (\ref{eq:V}). This can be pictured as a discrete change of the Hilbert space graph as $l$ is increased (keeping $L$ constant), where for each increase in $l$ the boundaries become more and more dominant up to the half-filling.
Finally, we note that the graph structure of $H_{\rm XX}$ in general corresponds to the \emph{Schreier graph} associated with the permutation group acting on the spins \cite{Bollobas_2013_book,Decamp_2020_PRR, Decamp_2020_PRR_0} which holds for arbitrary filling fraction.

\end{document}